\documentclass[preprint]{aastex}
\usepackage{natbib}

\shorttitle{A 7~MK Flux Rope}
\shortauthors{Aparna and Tripathi}
\begin{document}
\title{A 7~MK hot Flux Rope Observed by SDO/AIA}
\author{V. Aparna and Durgesh Tripathi}
\affil{Inter-University Centre for Astronomy \& Astrophysics, Post Bag - 4, Ganeshkhind, Pune 411007, India }

\begin{abstract}
A filament eruption was observed on October 31, 2010 in the images recorded by
the Atmospheric Imaging Assembly (AIA) on board the Solar Dynamic Observatory
(SDO) in its Extreme Ultra-Violet (EUV) channels. The filament showed a slow rise
phase followed by a fast rise and was classified to be an asymmetric eruption. In
addition, multiple localized brightening which was spatially and temporally
associated with the slow rise phase were identified leading us to believe
that the tether-cutting mechanism to be the cause of the initiation of the
eruption. An associated flux rope was detected in high temperature channels of AIA
namely 94~{\AA}  and 131~{\AA} corresponding to 7 MK and 11 MK plasma respectively. In
addition, these channels are also sensitive to cooler plasma corresponding to ∼1-2 MK.
In this study we have applied the algorithm devised by Warren et al. (2012) to remove
cooler emission from the 94~{\AA} channel to deduce only the high temperature structure
of the flux rope and to study its temporal evolution. We found that the flux rope was
very clearly seen in clean 94~{\AA} channel image corresponding to Fe XVIII emission which
corresponds to a plasma at a temperature of 7 MK. This temperature matched well
with that obtained using DEM analysis. This study provides important constrains in
the modelling of the thermodynamic structure of the flux ropes in CMEs.	
\end{abstract}


\section{Introduction}

Flux ropes are considered to be an integral part of coronal mass ejections (CMEs) in almost all of the existing theories attempting to explain the initiation and evolution of CMEs, though at different epochs during the CME evolution. Currently, there is great controversy on the subject as to when and how the flux ropes form. An account of prediction of epochs of formation of flux ropes in various models is summarised in \citet{2009A&A...498..295T}. A complete understanding of the nature of flux ropes and their evolutions will firstly lead to differentiate among the various theoretical models currently present and secondly have broad implications in the understanding of the physics of CME initiation. The latter is one of the most important questions in the field of solar physics, given its importance and implications in space weather and geo-space climate.

Flux rope like structures have commonly been observed in coronagraphic observations of CMEs recorded using the Large Angle and Spectrometric Coronagraph \citep[LASCO][; C2\&C3]{1995SoPh..162..357B} on board the Solar and Heliospheric Observatory (SoHO) and Cor-1\&2 \citep{2008SSRv..136...67H} on board Solar Terrestrial Relations Observatory (STEREO) spacecrafts \citep{2008SSRv..136..549C}. However, these observations have only confirmed the eventual presence of flux rope in CMEs without shedding substantial light to their origin. Therefore, low coronal observations are required. In off limb observations, flux ropes have been identified as large scale cavities with low density and high temperatures \citep{2006JGRA..11112103G, 2010ApJ...724.1133G, 2012ApJ...757...73K} and showing helical motion of plasma \citep[see e.g.,][]{2009ApJ...700L..96S}. These cavities are often related to the large scale polar crown filaments, where the filament material is located at the bottom of the flux rope and finally appear as a three part structure CME in coronagraphic images \citep{2004A&A...422..307C}.

The observations of flux ropes on disk have been difficult due to strong line of sight effects. Sigmoids are considered to be one of the best proxies for flux ropes on the solar disk \citep{2006SSRv..124..131G, 2009A&A...498..295T, 2009ApJ...698L..27T}. Studying the formation and evolution of sigmoids have given strong evidence towards the pre-existence of flux rope structures \citep{2009ApJ...698L..27T}. \citet{2009A&A...498..295T} compared the observables predicted by a pre-existing flux rope model with observations for 6 well observed events, and concluded that these observations could be explained well with the model with pre-existing flux rope. However, direct observations of a flux rope from its birth has not been possible till the launch of the Solar Dynamics Observatory (SDO) in 2010.

Almost all the models proposed for the formation of flux rope involves magnetic reconnection, that will lead to emission from hot plasmas. Since all the EUV channels in which the Atmospheric Imaging Assembly (AIA, \citealp{2012SoPh..275...17L}) observes are sensitive to plasmas at different temperatures, these observations could be used as a thermometer for coronal plasmas. However, it is non-trivial to interpret the AIA observations directly as temperature diagnostics as AIA channels often have contributions from multiple emission lines sensitive to different temperature plasmas \citep{2010A&A...521A..21O}. Using the AIA observations of a flaring active region, \citet{2011ApJ...727L..52R} have constrained the temperature of flux ropes to between 5~MK and 18~MK suggesting that the flux ropes could reach flare-like temperatures. In this observation, the flux rope was only observed in 94 and 131~{\AA} channels, which are primarily sensitive to 7~\&~11~MK and was not present either in 193~{\AA} or in 171~{\AA}. \citet{2015ApJ...808..117N} made a survey to establish whether flux ropes commonly exist in CME eruptions in AIA data taken using 131, 171 and 304~{\AA} channels. About 32\% of their eruptive events out of 141 showed features of flux-rope like structure in 131~{\AA} and none in the rest of the wavelengths, again suggesting high temperatures of flux ropes. Similarly, \cite{2015A&A...580A...2Z} conducted a large survey that included on-disk as well as limb observations of 1354 "flux-rope proxies". For about 50\% of their events, flux-rope proxies were also seen in 171 and 304~{\AA}. The authors attribute the structures seen in the cooler channels to secondary flux ropes that form due to reconnection of the coronal arcade and lifting of the cool material higher into the atmosphere during the process \citep{2010SSRv..151..333M}. 

In addition to directly associating a temperature to the structures seen in different AIA channels, which could at times be non-trivial and might be misleading due to contributions from multiple emission lines sensitive to different temperature plasmas, Differential Emission Measure (DEM) inversion methods have also been adopted \citep[see e.g.,][]{2011ApJ...727L..52R, 2012ApJ...761...62C, 2013ApJ...778..142T, 2013ApJ...764..125P}. DEM analysis of various regions of the CME structures done by \citet{2012ApJ...761...62C} have shown a temperature \textgreater\ 8 MK for the flux ropes. \citet{2013ApJ...778..142T} analysed a partially erupting prominence using the DEM technique and speculated the presence of a flux rope in the temperature range from 5{--}10~MK. The temperature map of a failed eruption observed by \citet{2013ApJ...764..125P} showed the flux rope structure measuring up to 10~MK. Although, all these above mentioned studies provide a range of temperature of the flux ropes at a given instance of time, they do not provide the details of the evolution of high temperature plasma. This is most likely due to the fact that pixel-wise DEM inversion for narrow band images is a cumbersome and involved process and computationally intensive. Therefore, it may become impractical to perform such a study. 

The main aim of this paper is to study the initiation and thermodynamic evolution of the filament eruption and associated flux rope that was observed on October 31, 2010. To this end we study the kinematics of an erupting filament and associated flux rope and determine the temperature of the flux rope. In addition, we also study how the hot plasma in the flux rope evolves with time during eruption. We have used a novel technique recommended by \citet{2012ApJ...759..141W} to remove the cool temperature component from the 94~{\AA} channel observed by the AIA. The use of this method provides a unique opportunity to study the evolution of true high temperature plasma in the flux rope, which has not been possible so far, due to the reasons mentioned above. The remainder of the paper is structured as follows. In Section~\ref{obs} we present the observations and data analysis. We discuss the temperature structure of the associated flux rope in Section~\ref{temp}. Finally, we summarise the results and conclude in Section~\ref{discussion}.

\section{Observations and Analysis}~\label{obs}

A filament eruption from a small active region situated in the northern hemisphere of the Sun on Oct 31, 2010 was recorded by AIA on board SDO. AIA observes the Sun in nine channels namely 94, 131, 171, 193, 211, 304, 335, 1600 and 1700~{\AA}. The first six channels primarily correspond to different ionization states of Iron. Although, only one wavelength is assigned to each channel, multiple  spectral lines contribute to the whole emission in the respective solar images. For example, the 94~{\AA} channel have contributions from \ion{Fe}{18} at ~93.93~{\AA} line as well as emission due to \ion{Fe}{10} forming at a wavelength of 94.01~{\AA} among other weaker lines. Similar is the case for 131~{\AA} as well as the other channels. For more information see \cite{2010A&A...521A..21O}, \cite{2011A&A...535A..46D}, \cite{2012SoPh..275...17L} and \cite{2012SoPh..275...41B} regarding the AIA instrument and the responses of various channels due to different kind of plasma. 

Since the aim of this work is to study the high-temperature structure of flux ropes we concentrate on the high temperature channels of AIA, namely the 94~{\AA} and 131~{\AA} channels. However, as mentioned earlier, these channels are contaminated with warm emissions coming from the corona corresponding to various loop structures as well as the diffuse emission which peaks around $\log\, T =  6.25$ which is characteristic of 193~{\AA} channel. By doing a comparison of various observations recorded by 94~{\AA} along with that of 171~{\AA} and 193~{\AA} channels, \cite{2012ApJ...759..141W} devised a scheme to obtain the emission due to pure \ion{Fe}{18} line, which corresponds to plasma at a temperature of $\sim$~7~MK.

Therefore, in this study, we have applied the algorithm developed by \cite{2012ApJ...759..141W} to the filament eruption in order to separate the hot components from the cooler ones. In order to estimate the higher temperature component in the 94~{\AA}, \citeauthor{2012ApJ...759..141W} obtained the following scheme:

\begin{equation}
I_\textrm{94warm} = 0.39 \sum_{i=0}^{3} a_i \Big[\frac{fI_\textrm{171} + (1-f)I_\textrm{193}}{116.54}\Big]^i
\end{equation}

The value of f is taken as 0.31 such that the estimated intensities correlate well with the observed intensities. The coefficients of the polynomial fit are $-7.31 $x $10^\textrm{-2}$, $9.75$ x $10^\textrm{-1}$, $9.9$ x $10^\textrm{-2}$ and $-2.84$ x $10^\textrm{-3}$. The constants are the scaling factors derived from the median intensities. For more details see \cite{2012ApJ...759..141W}. We have also compared our results with that obtained using \cite{2013A&A...558A..73D} algorithm for the removal of the lower temperature contribution in the 94~{\AA} channel.

\subsection{Overview of the Eruption}

\begin{figure}
\includegraphics[scale=0.7,trim={3.5cm 0cm 3.5cm 0cm},clip=true]{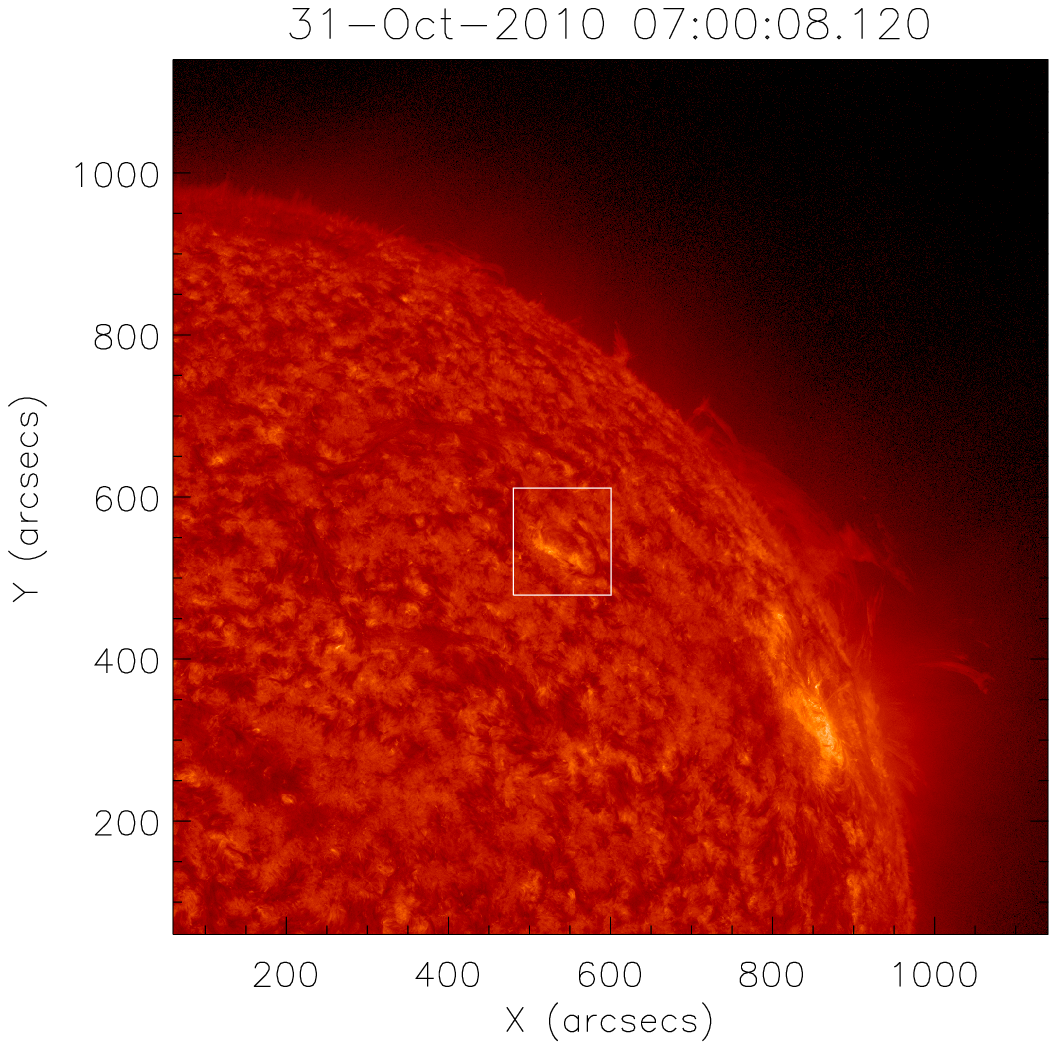}
\includegraphics[scale=0.7,trim={3.0cm 0cm 3.0cm 0cm},clip=true]{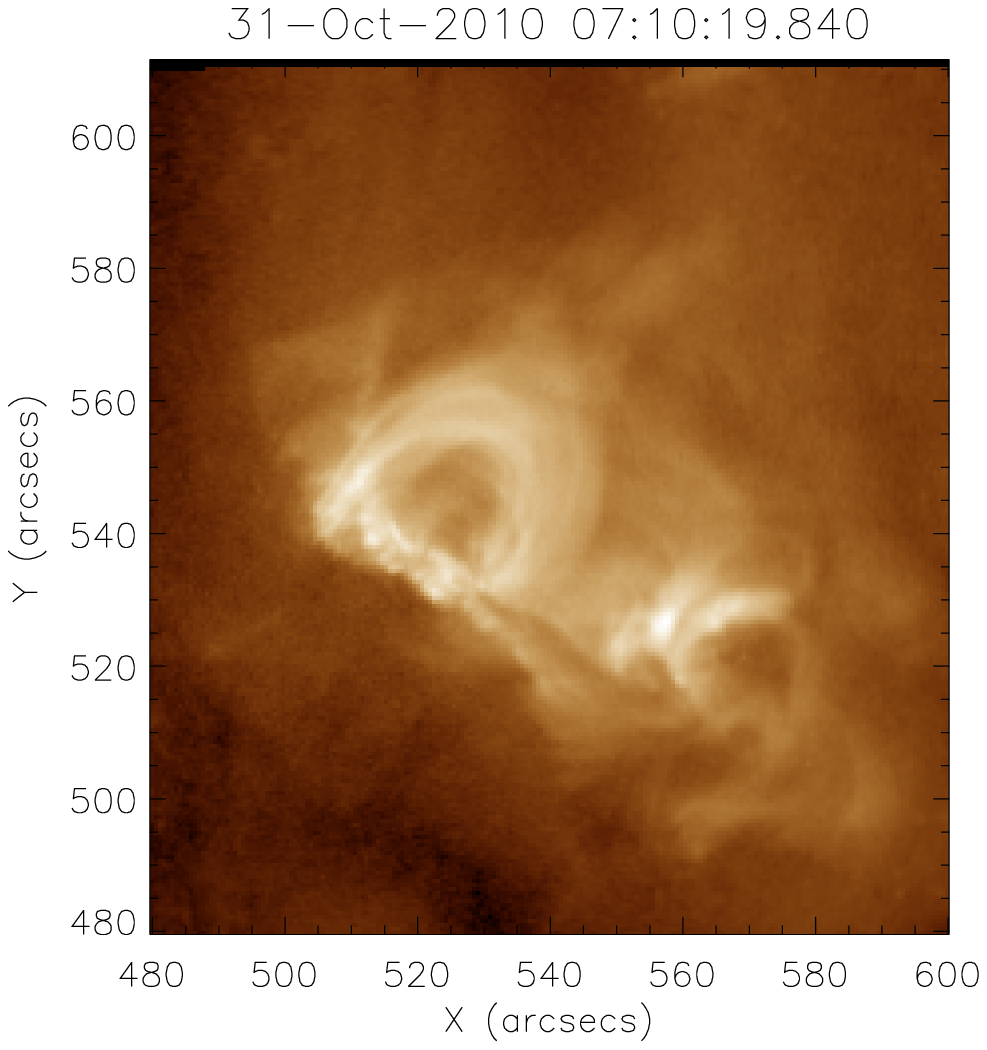}
\caption{The northwestern hemisphere of the Sun on Oct 31, 2010 (left) shows the 
region of interest marked by a white bounding box. The region of interest is shown 
in 193~{\AA} to the right of it.\label{context}}
\end{figure}

\begin{figure}
\includegraphics[scale=0.6,trim={4.7cm 1.4cm 3.2cm 0.6cm},clip=true]{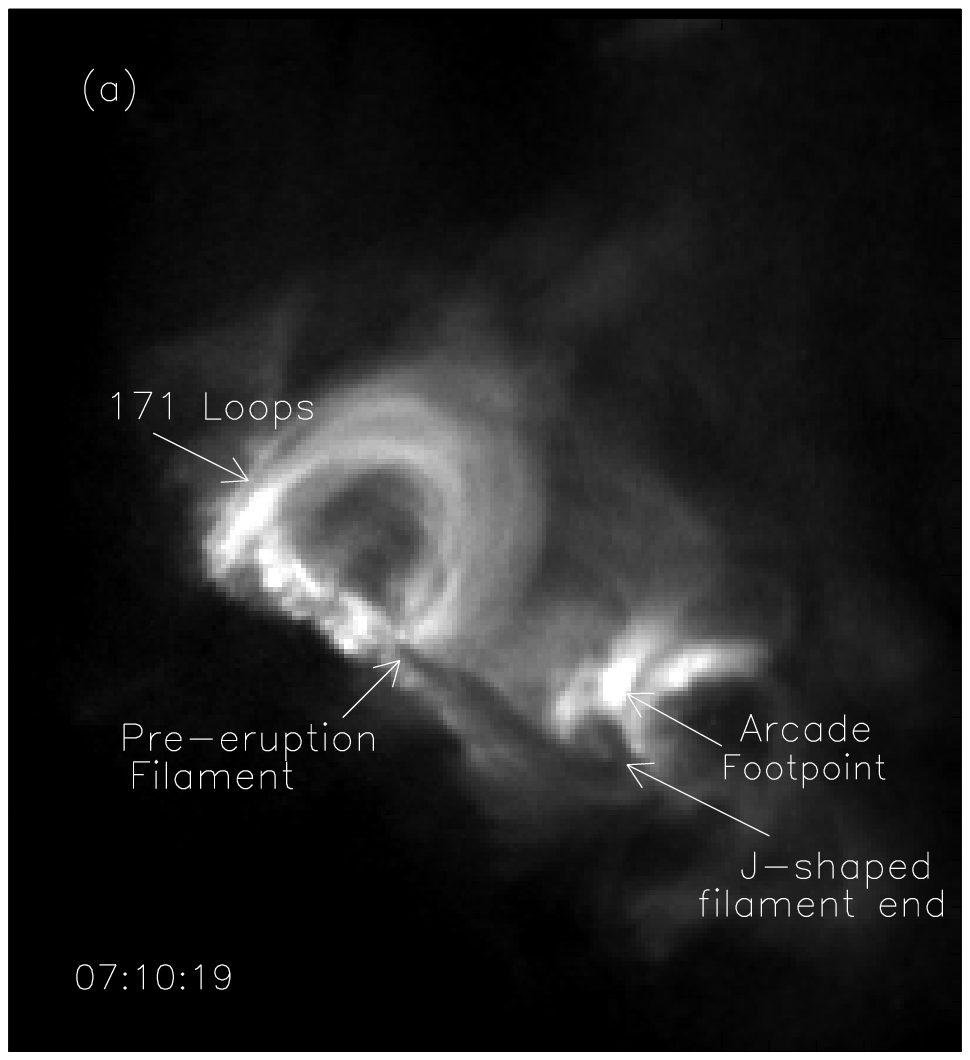}\includegraphics[scale=0.6,trim={4.7cm 1.4cm 3.2cm 0.6cm}, clip=true]{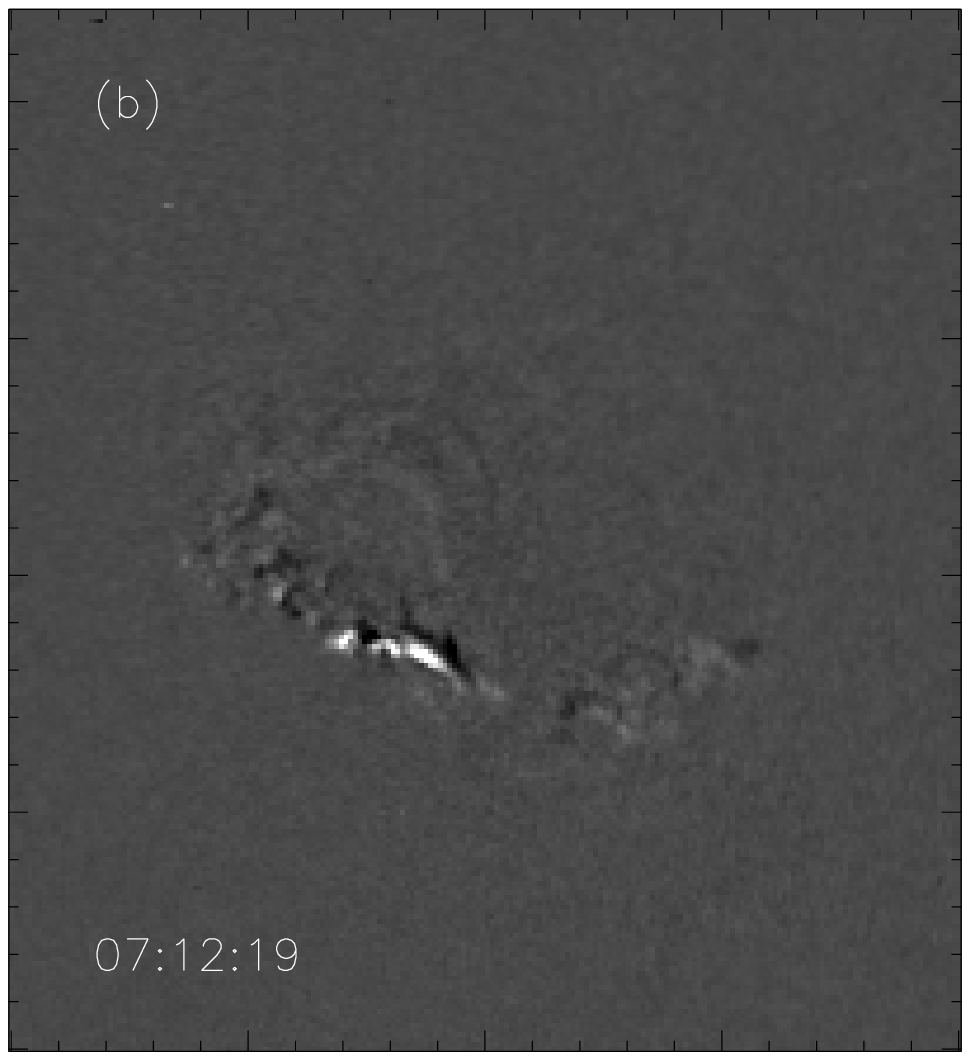}\includegraphics[scale=0.6,trim={4.7cm 1.4cm 3.2cm 0.6cm},clip=true]{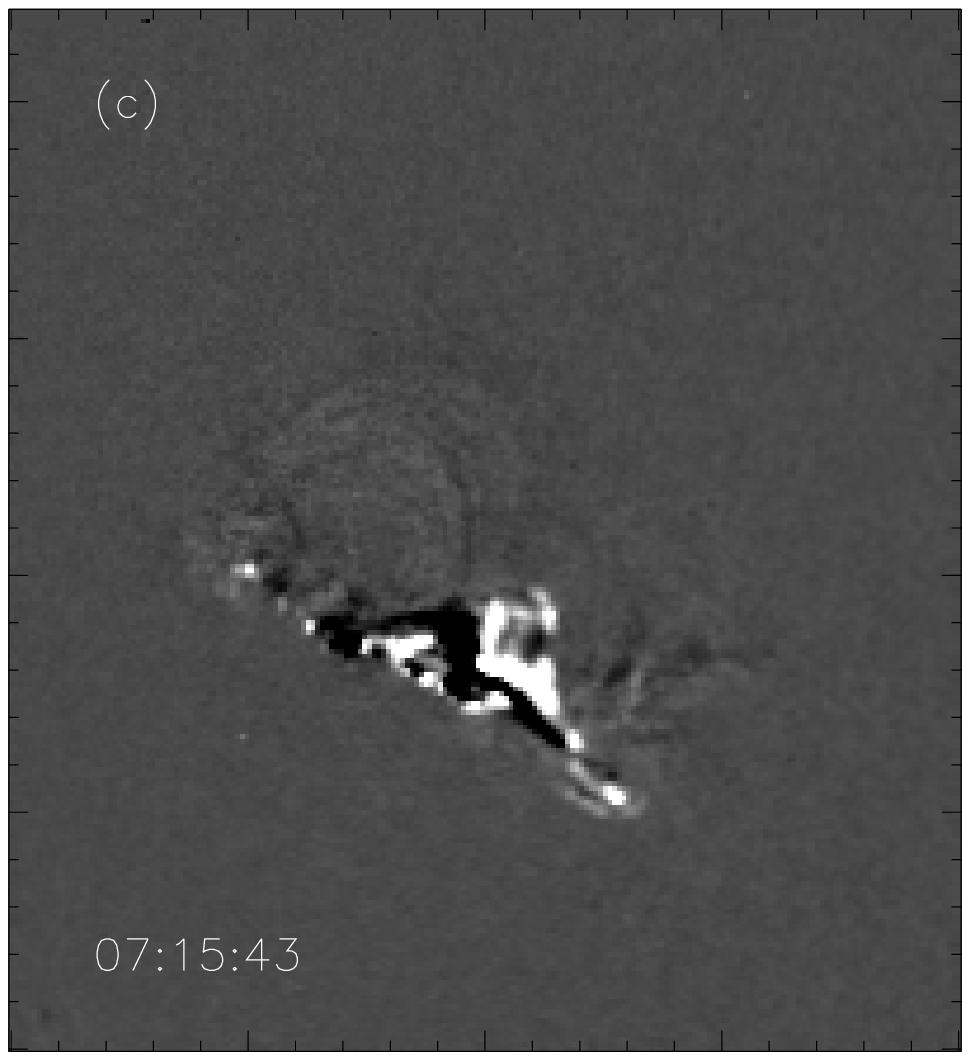}

\includegraphics[scale=0.6,trim={4.7cm 1.4cm 3.2cm 0.6cm},clip=true]{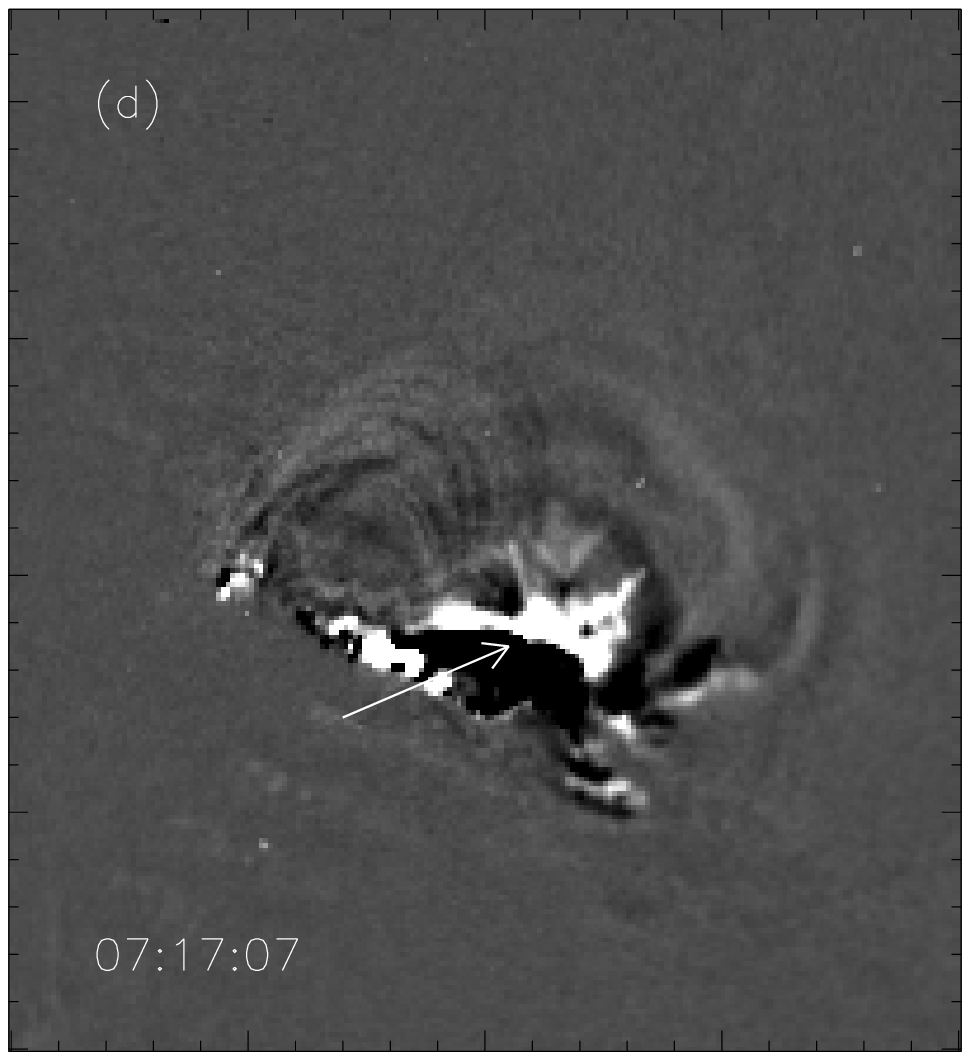}\includegraphics[scale=0.6,trim={4.7cm 1.4cm 3.2cm 0.6cm}, clip=true]{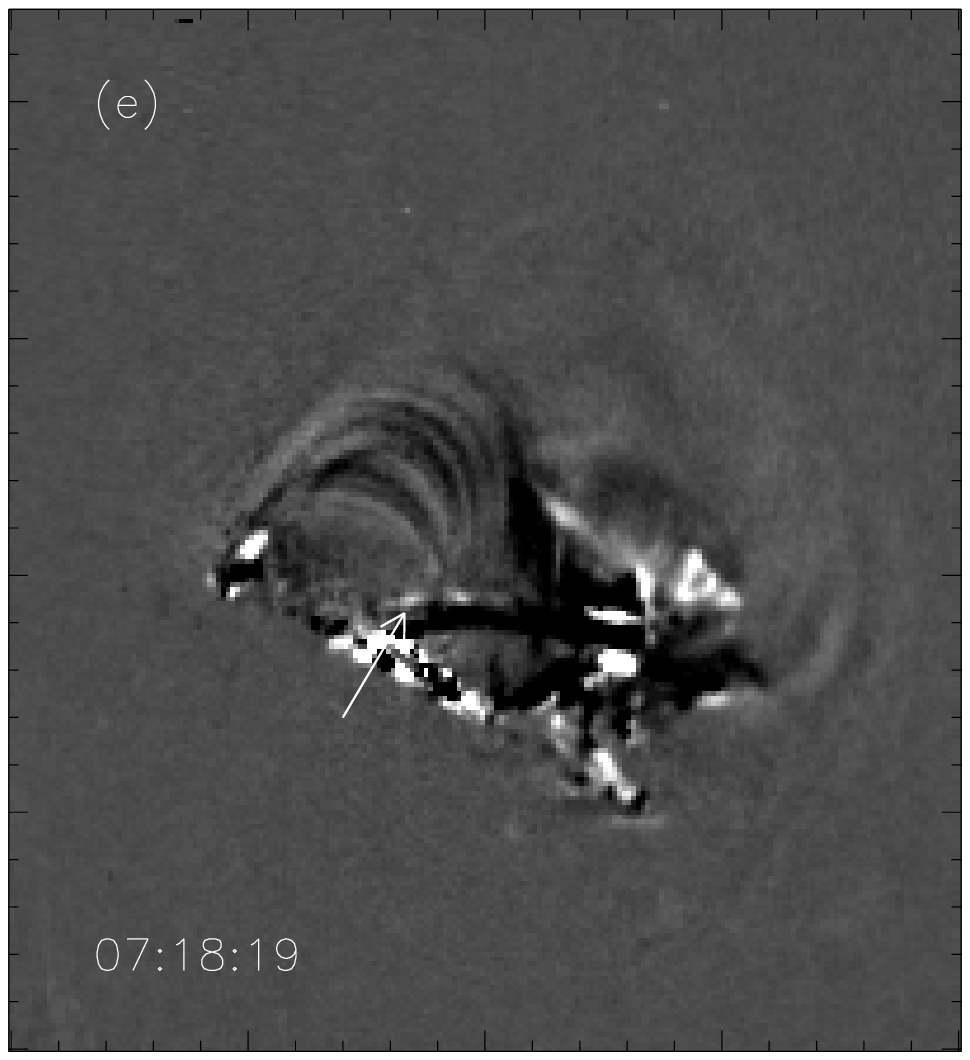}\includegraphics[scale=0.6,trim={4.7cm 1.4cm 3.2cm 0.6cm},clip=true]{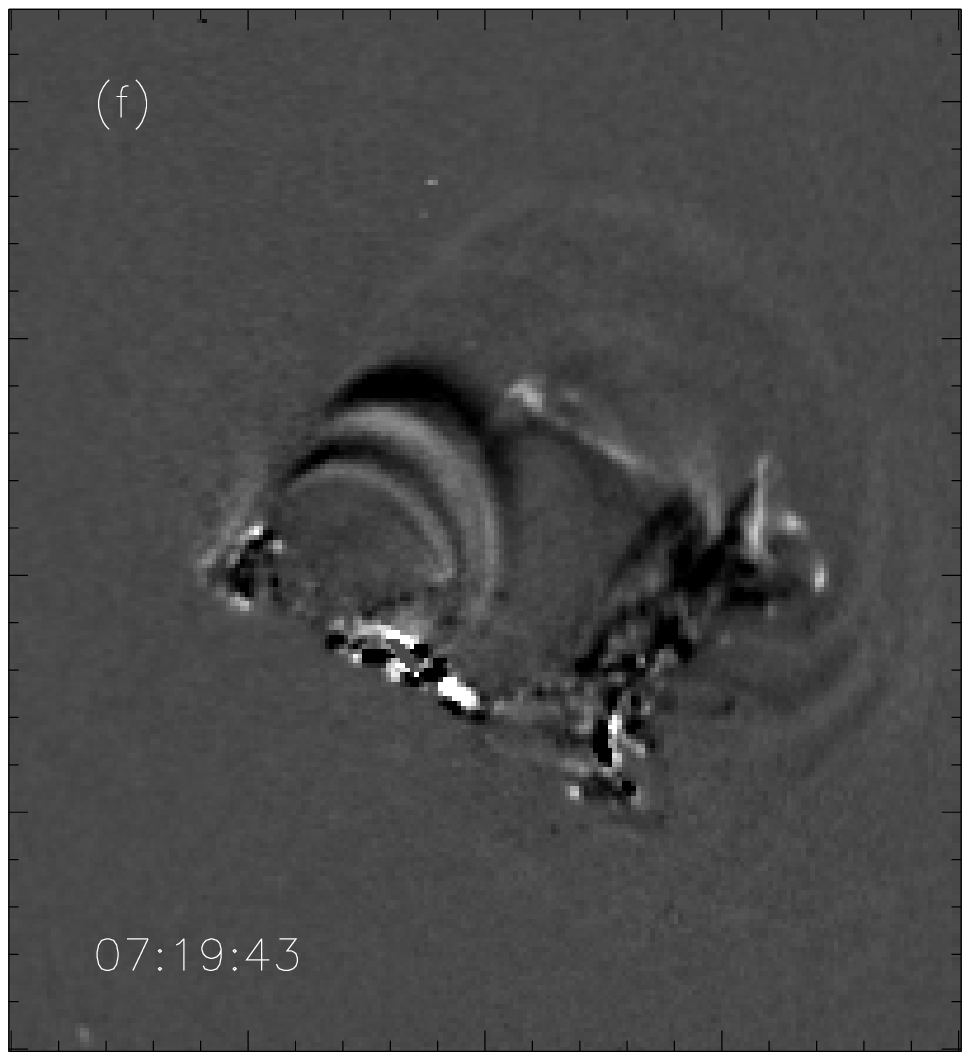}

\includegraphics[scale=0.6,trim={4.7cm 1.4cm 3.2cm 0.6cm},clip=true]{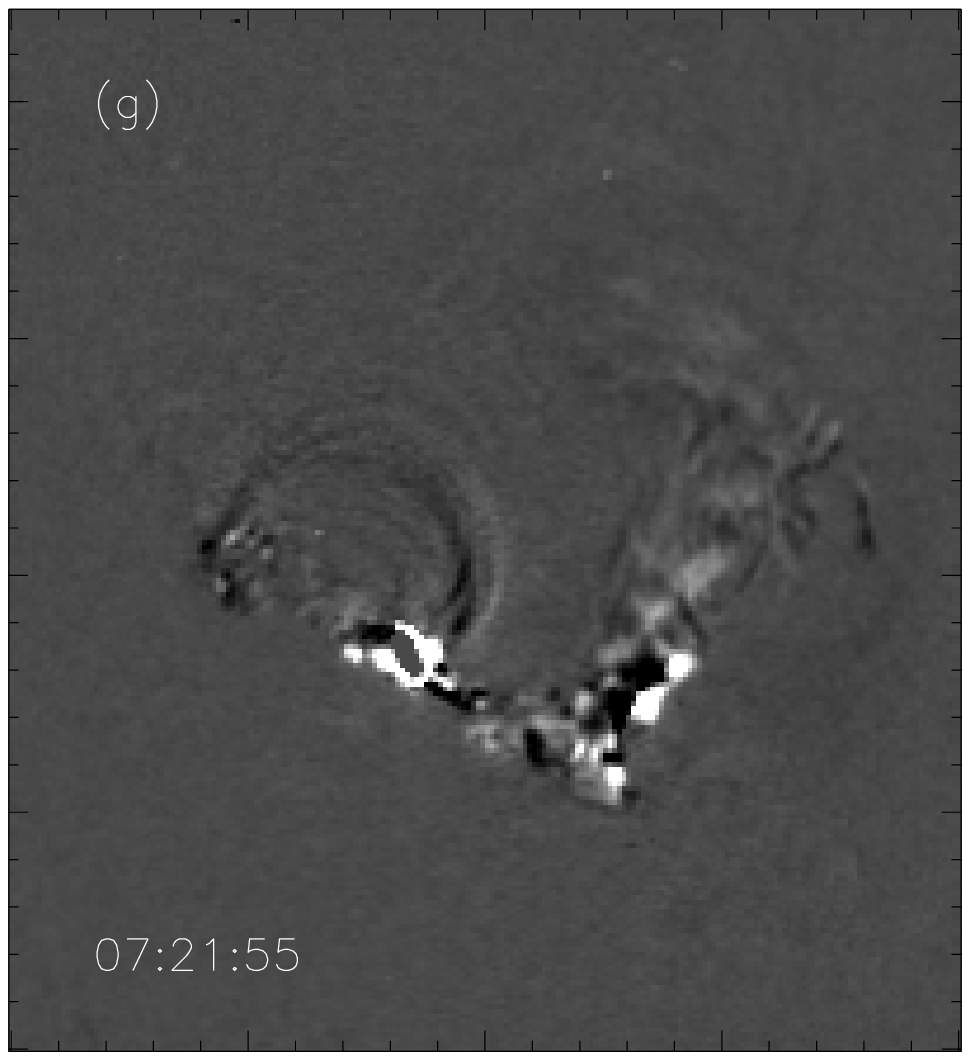}\includegraphics[scale=0.6,trim={4.7cm 1.4cm 3.2cm 0.6cm}, clip=true]{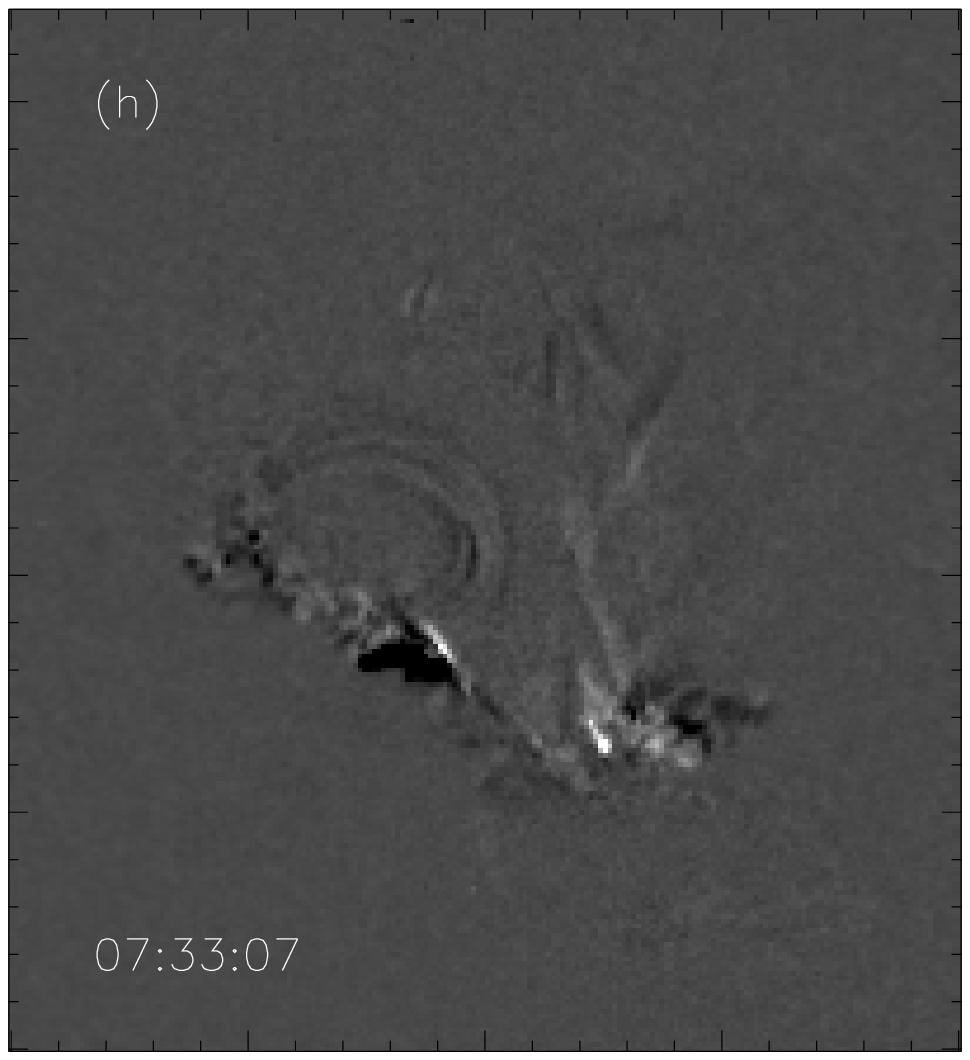}\includegraphics[scale=0.6,trim={4.7cm 1.4cm 3.2cm 0.6cm},clip=true]{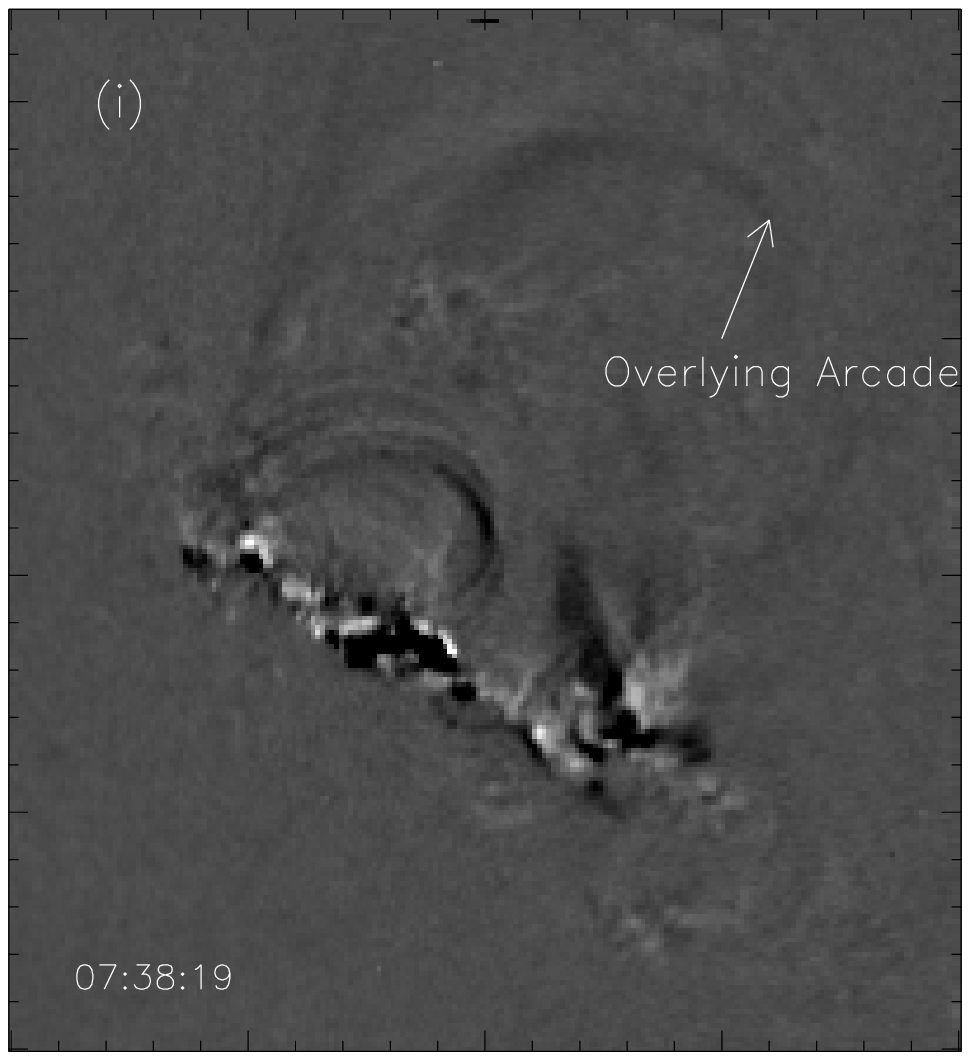}

\caption{A sequence of running-difference images using 193{\AA} showing the evolution of the 
erupting filament, including the untwisting motion. The arrows in figures d and e point to the location of the cutting of the filament threads.} An animation is available in the online version of the paper.\label{evolution}
\end{figure}

The filament eruption occurs between 07:11 and 07:35 UT and are observed in all the EUV channels of AIA. Figure~\ref{context} displays the region of interest on the solar disk in 304~{\AA} (left panel) and a blow-up of the boxed region in the right panel in 193~{\AA}. The filament eruption begins at 07:11~UT from the coordinates [522, 528] arcsec, where intense brightening is seen in 94 and 131~{\AA} at 07:11 UT. Figure~\ref{evolution} displays the sequence of images showing the evolution of the eruption. Running difference images are obtained using 193~{\AA} by subtracting an average of three images prior to every image \citep{2014ApJ...797..131S}. Different structures are labelled by arrows in different panels shown in Figure~\ref{evolution}.

\begin{figure}
\includegraphics[scale=0.45,trim={2.7cm 0.3cm 1cm 0.2cm},clip=true]{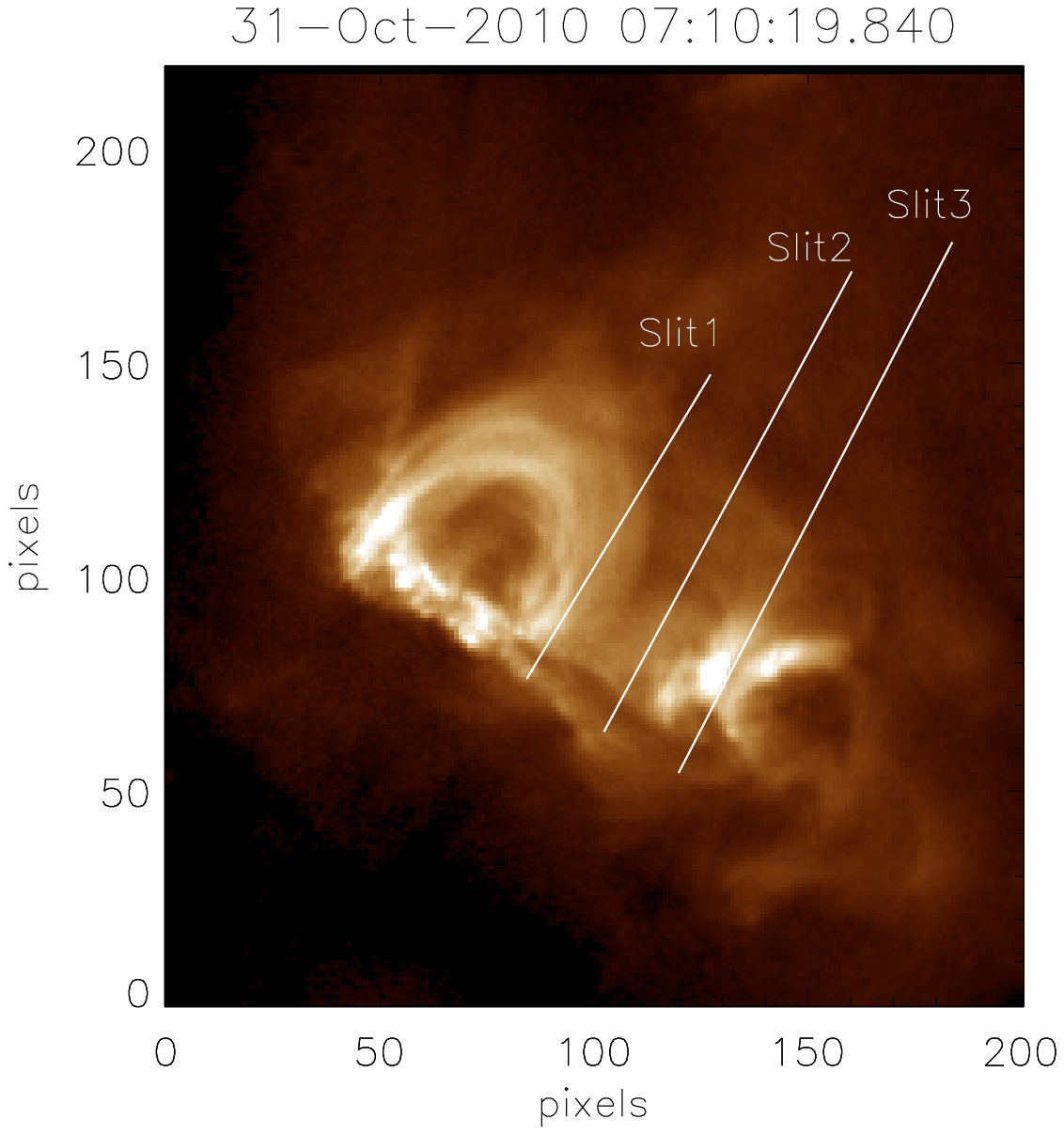}\includegraphics[scale=0.6]{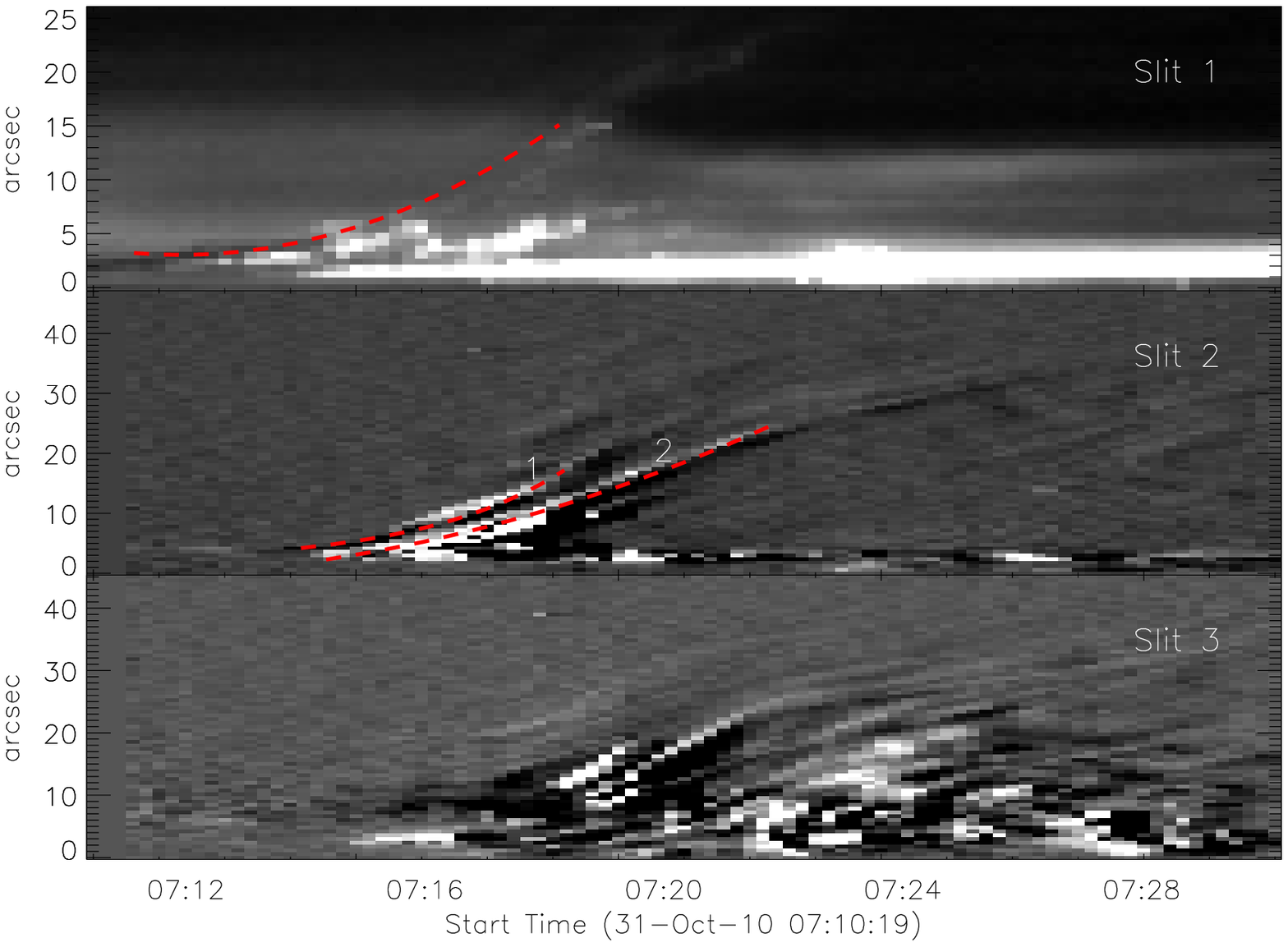}
\caption{Time distance plots of the erupting filament fot the selected slit positions (left) is shown on the right. The slits track the asymmetric eruption of the filament. The dashed red lines on the time distance plots represent the fitting. \label{TSD}}
\end{figure}

In order to study the kinematics of the erupting filament, we have created time slice diagrams (TSD) using 193~{\AA} images using three artificial slits across the erupting filament. The left panel in Figure~\ref{TSD} displays the three slits plotted over the AIA image. The right panel shows the TSD obtained for the three slits as labelled. Note that the TSD are obtained using the running difference images. The eruption is slow in the beginning for about 4 minutes and shows accelerating features afterwards. The slow-rise phase of the eruption lasts until about 07:15~UT and the fast-rise occurs between 07:15 and 07:27 UT after which the filament material falls back to the surface towards the south-western leg and is seen until 07:42 UT. The TSD for slit-1 could be fitted with a second order polynomial and we obtain an acceleration of $35.59\ km/s^2 $. TSD for slit-2 could be best fitted with an exponential curve (curve 1) and a second order polynomial (curve 2, acceleration = $23.95\ km/s^2$). The slow rise in the early phase of the eruption followed by an accelerated phase is apparent. The falling of the material along the south-western footpoint in the later phase of the eruption is apparent in the bottom panel of Figure~\ref{TSD}. The TSDs show that the eruption is asymmetric, which starts from the north-eastern footpoint and propagates towards the south-west, as was first described by \citet{2006A&A...453.1111T}.	

In the cooler AIA channels (304, 171 and 193~{\AA}), the filament is seen as a dark absorption feature during the initial rise phase (Figure~\ref{evolution}~a~\&~b). The western end of the filament is curved like a 'J' and the eastern end is rooted near the brightening region from where the eruption begins. The overlying arcade structure (Martin 1998) engulfing the filament has one of its foot near the 'J' end of the filament and the other foot in front of the eastern end of the '171 loops' (see Figure 2a \& b). The arcade (marked in panel i of Figure 2) is pushed higher as the filament erupts. This is seen from 07:17:36 UT until the end of the observations. It expands up to a height of 620\arcsec in the images. Although the structure is very diffuse at this height, it is best seen in 193~{\AA} (Figure 3g, h \& i). As the filament eruption is in progress, several threads that make the filament get cut. This cutting resembles the tether-cutting as described in \cite{2001ApJ...552..833M, 2006A&A...458..965C, 2007A&A...472..967C, 2013ApJ...778L..36L}. These tether-cutting events occur between 07:15:00-07:15:12, 07:17:00-07:17:12, 07:17:48-07:18:00, 07:18:07-07:18:21 and 07:20:36-07:21:24 UT. One example can be seen when Figures 2d and 2e are compared. The filament material does not get ejected but falls to the surface, some of the threads get tethered from the large arcade structure causing the arcade to rise further higher (seen between 07:29:48-07:31:48 UT). Blob shaped brightening, localised in space and time, are seen at the locations where the threads are cut in all these instances of tether cutting. 

In the hotter channels of AIA such as (94~{\AA} and 131{\AA}) bright emission is seen prior to the beginning of the filament eruption and continues to brighten along the filament length thereafter. The untwisting of the filament as it rises, the expansion of the overlying arcade and the foot-point of the arcade are seen well in 171 and 335~{\AA}. The arcade, its foot and the loops behind the filament region (marked 171 loops in Figure 2a) are visible in the cooler channels except in 304~{\AA}.

\begin{figure}
\includegraphics[scale=0.55,trim={4.8cm 2cm 4.8cm 2cm},clip=true]{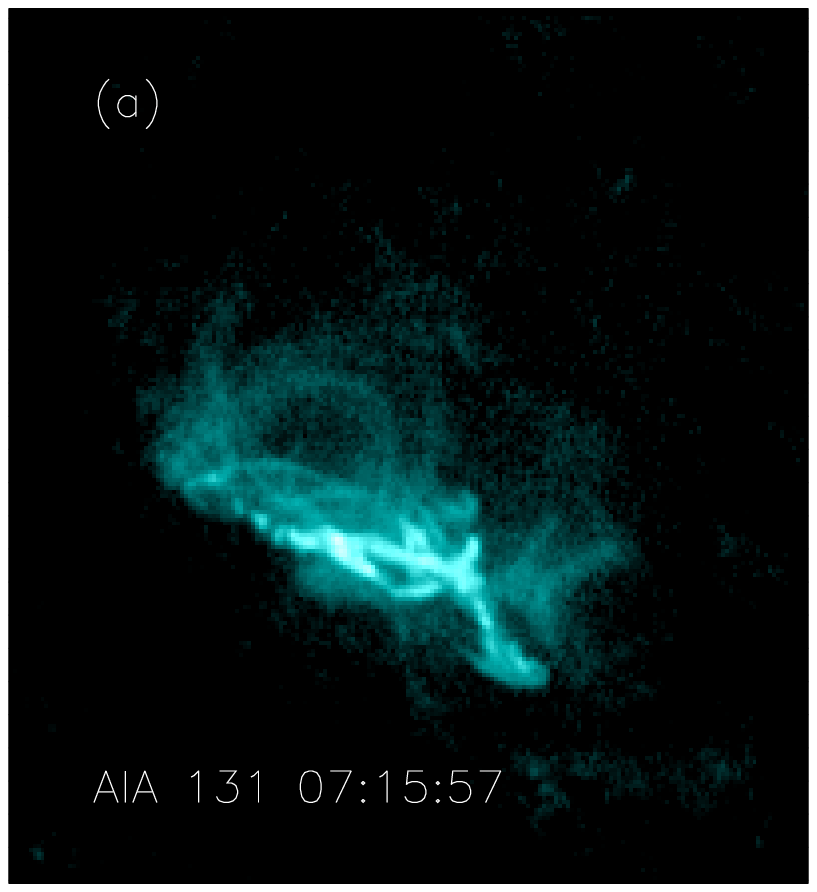}\includegraphics[scale=0.55,trim={4.8cm 2cm 4.8cm 2cm},clip=true]{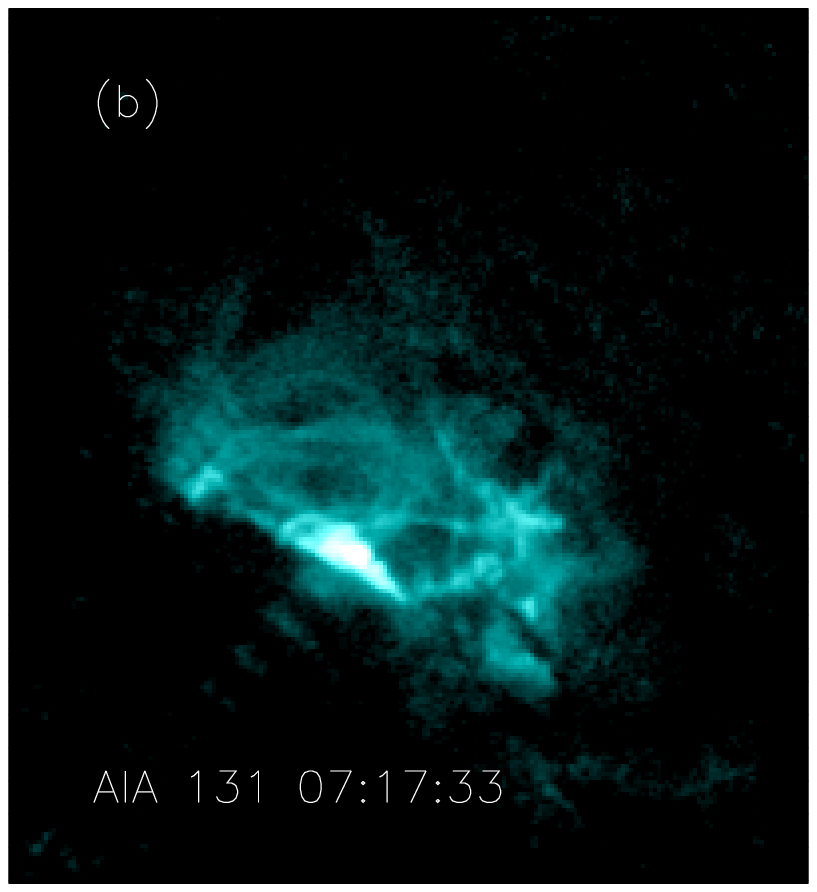}\includegraphics[scale=0.55,trim={4.8cm 2cm 4.8cm 2cm},clip=true]{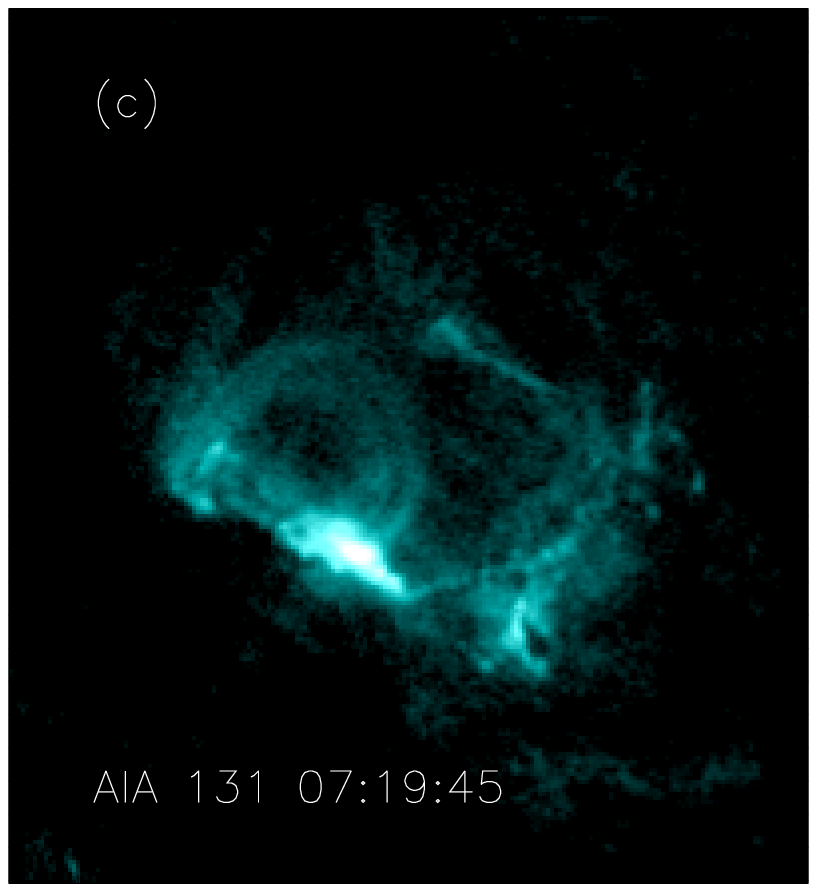}\includegraphics[scale=0.55,trim={4.8cm 2cm 4.8cm 2cm},clip=true]{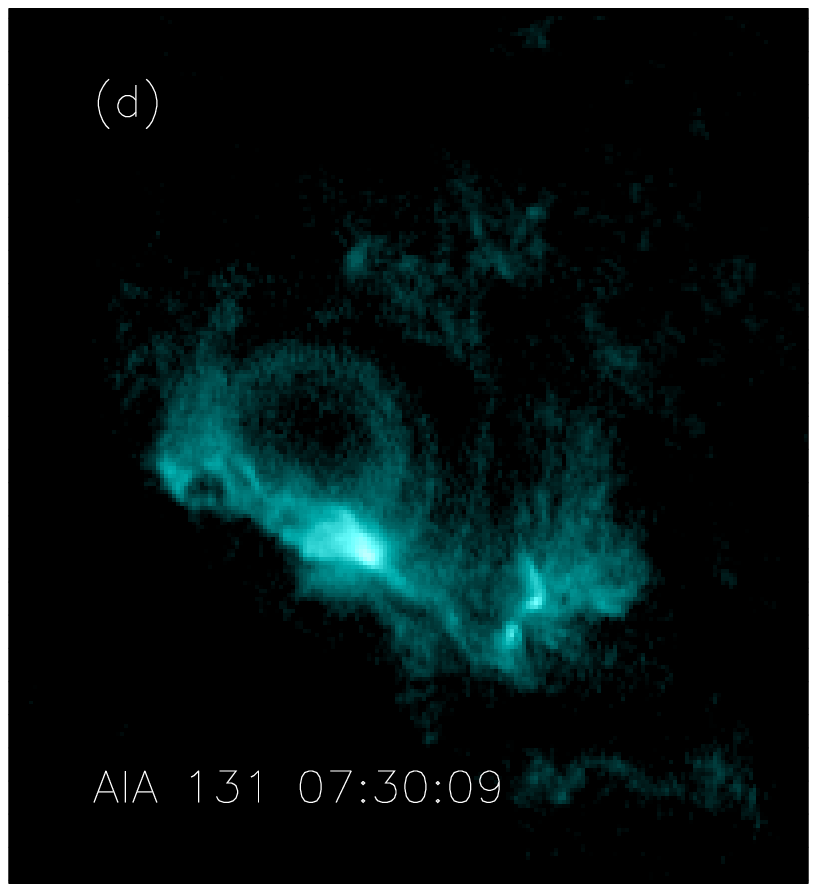}

\includegraphics[scale=0.55,trim={4.8cm 2cm 4.8cm 2cm},clip=true]{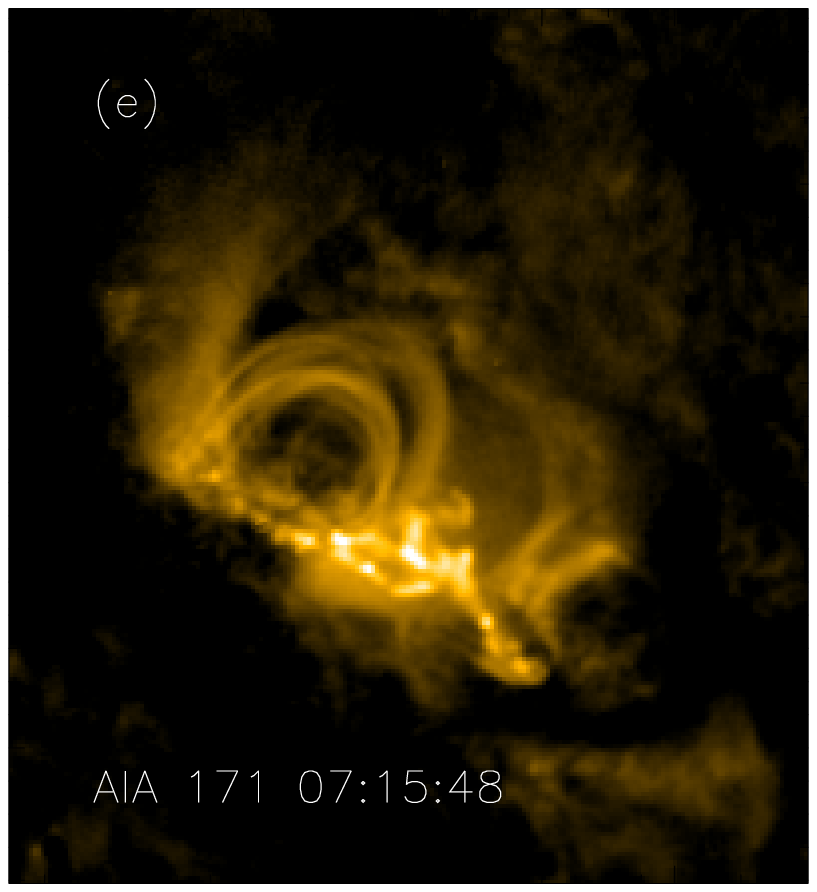}\includegraphics[scale=0.55,trim={4.8cm 2cm 4.8cm 2cm},clip=true]{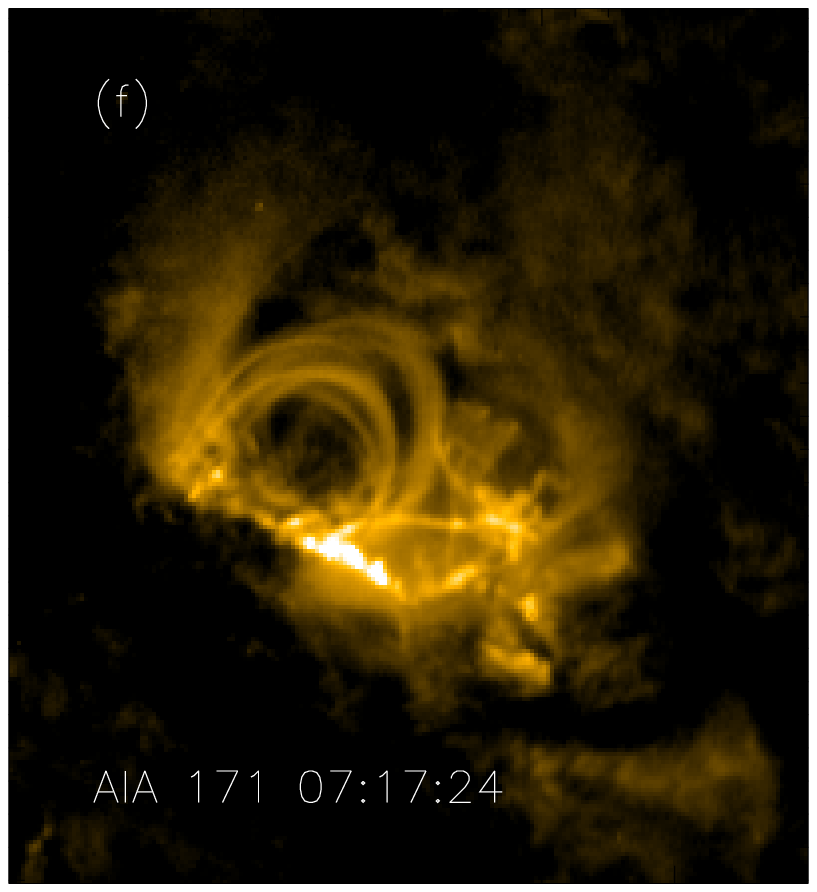}\includegraphics[scale=0.55,trim={4.8cm 2cm 4.8cm 2cm},clip=true]{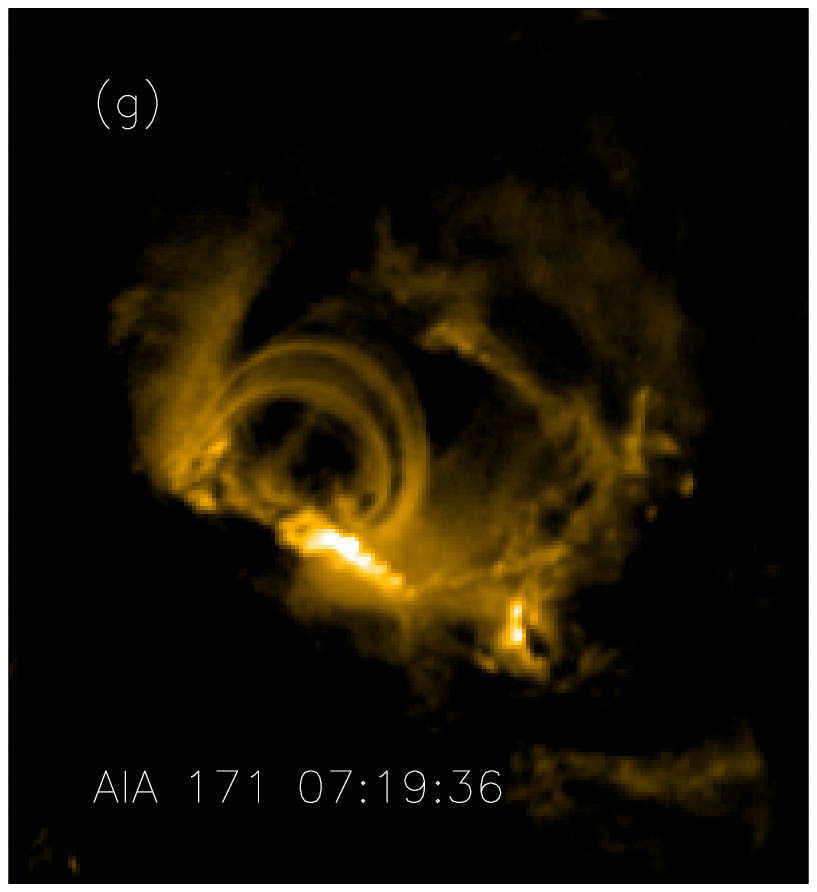}\includegraphics[scale=0.55,trim={4.8cm 2cm 4.8cm 2cm},clip=true]{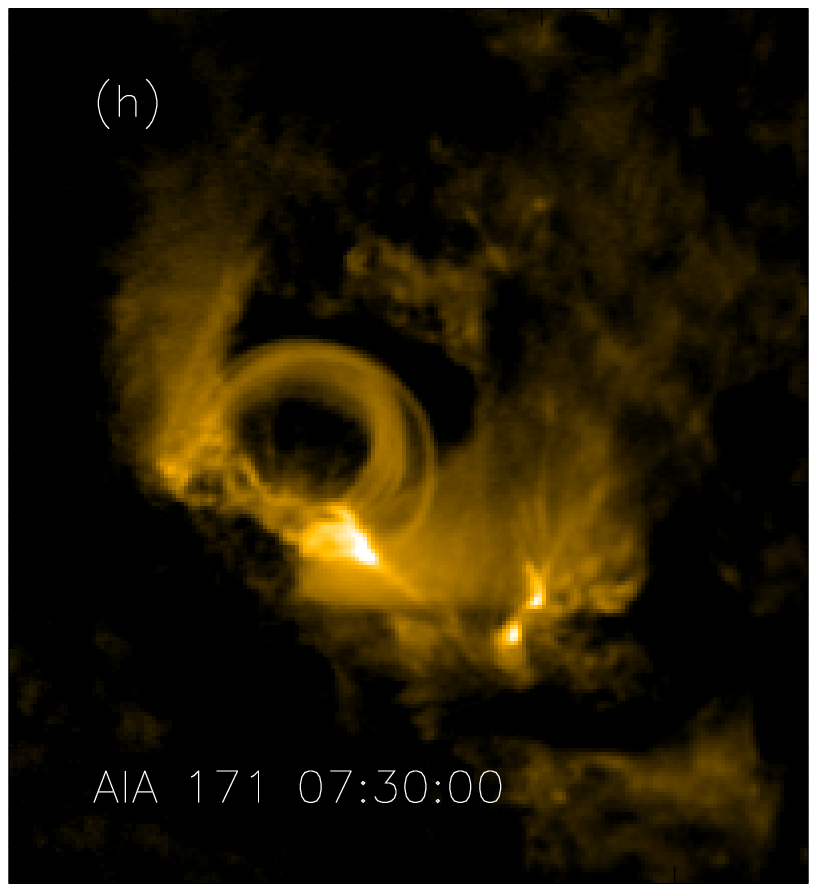}

\includegraphics[scale=0.55,trim={4.8cm 2cm 4.8cm 2cm},clip=true]{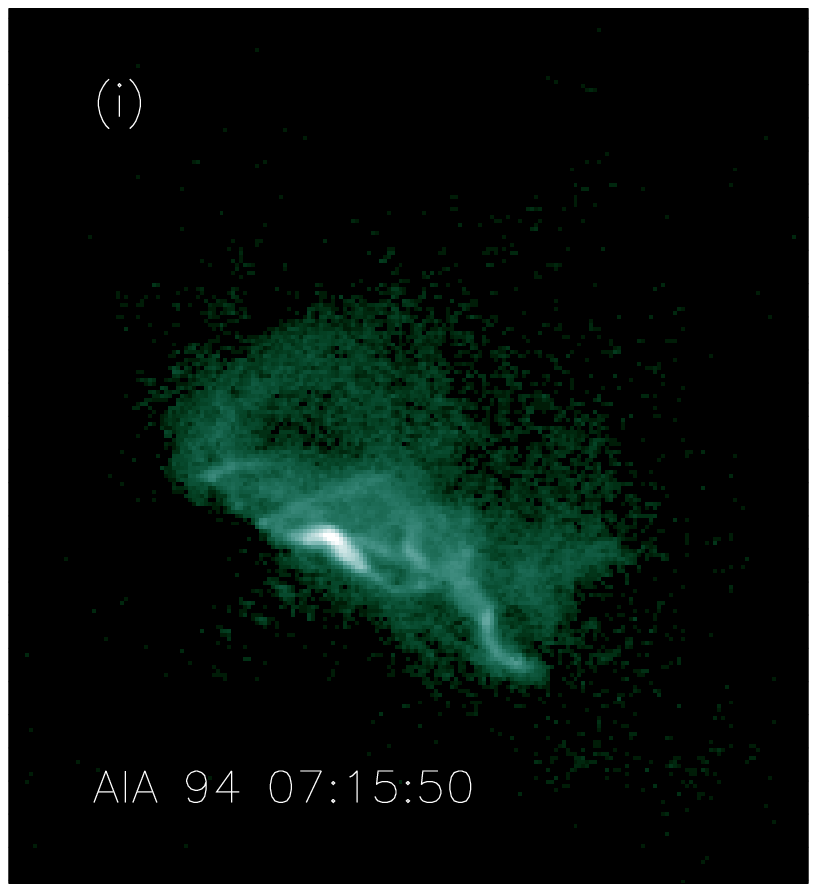}\includegraphics[scale=0.55,trim={4.8cm 2cm 4.8cm 2cm},clip=true]{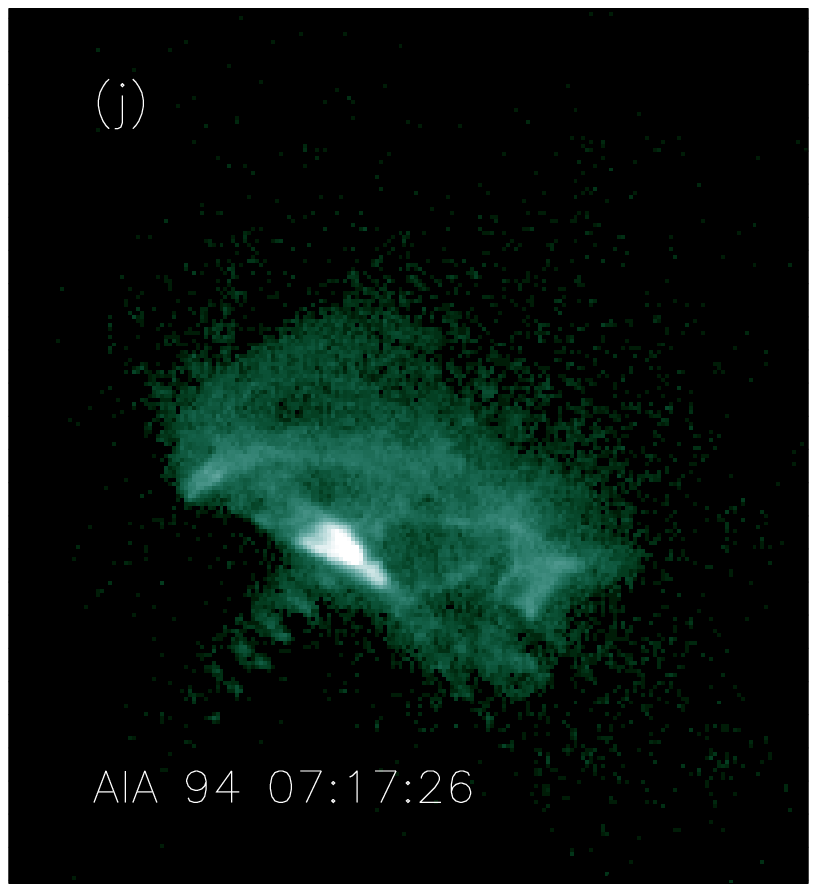}\includegraphics[scale=0.55,trim={4.8cm 2cm 4.8cm 2cm},clip=true]{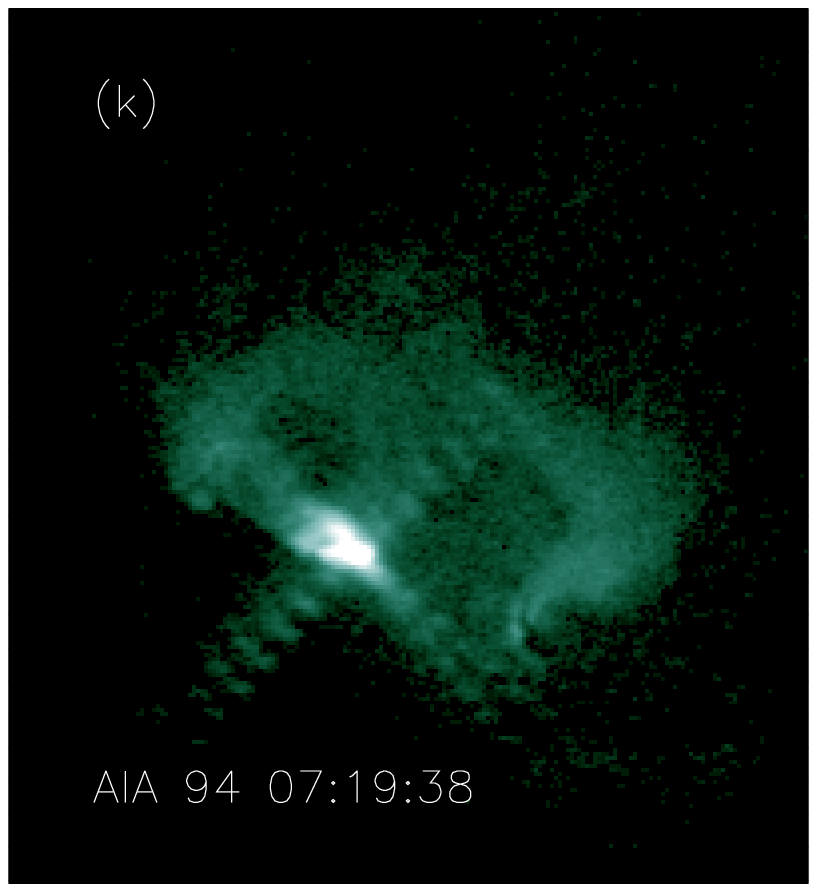}\includegraphics[scale=0.55,trim={4.8cm 2cm 4.8cm 2cm},clip=true]{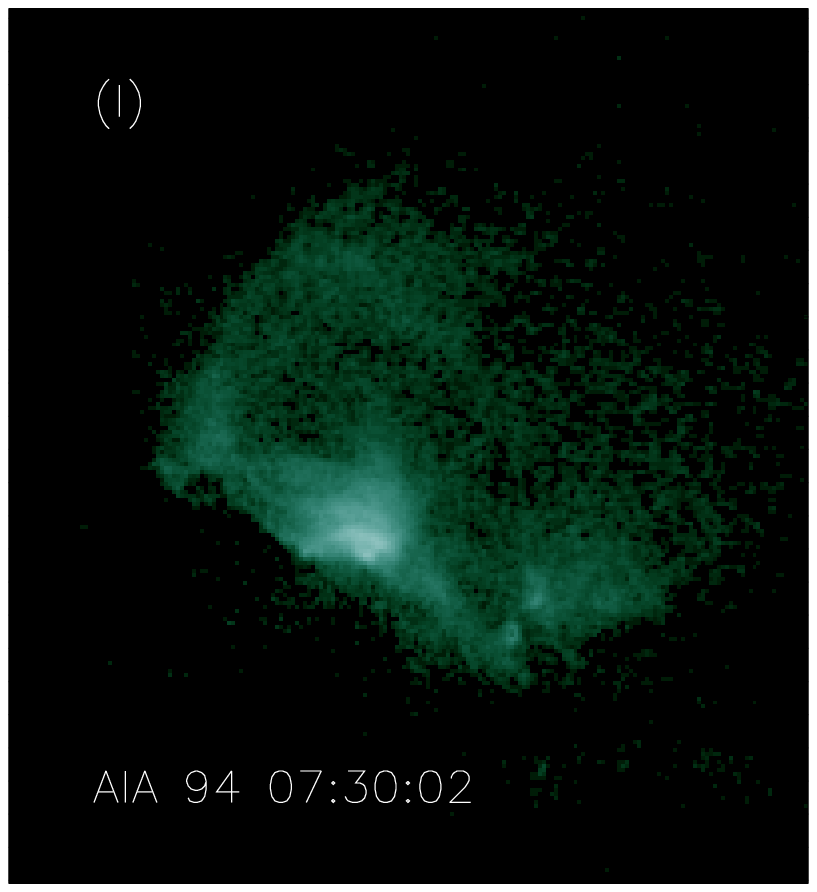}

\caption{AIA 131, 171 and 94~{\AA} images at different times during the filament eruption are shown in first, second and third rows, respectively. The flux rope structure is visible in 131~{\AA} and 94~{\AA} initially (figures a, b, i and j) and can be seen only in 94~{\AA} as the eruption progresses higher (as seen in k and l). \label{multi}}
\end{figure}

Figure~\ref{multi} shows selected frames in 131, 171 and 94~{\AA}. In 94 and 131~{\AA}, as the filament eruption is under progress, two new strands intertwined with the filament start to rise at 07:15:38~UT. Part of the left most strand is connected behind the west end of the '171 loops' and part of it along the filament structure (Figure~\ref{multi}~a\&~e). This is apparent as the two strands continue to rise as a single structure. These strands possibly make the structure of the features seen in the 94~{\AA} images shown in the bottom row in Figure~\ref{multi}. This structure itself is not visible in any of the cooler channels. However, intermittent brightening is seen near the foots of the two strands at around 07:16:12~UT in 171 and 304~{\AA}. The intertwined nature of the flux rope like structure is evident between 07:18:09 and 07:18:21~UT in 131~{\AA} when a thin thread connected near the root of the left most strand is cut from the twisted region in the central part. In 94~{\AA} and 131~{\AA}, the flux-rope is visible as a diffuse structure (Figure~\ref{multi}~b,~j,~k~\&~l) that rises along with the overlying arcade structure until the end of the observations. The overlying arcade is not seen in 131~{\AA} and is diffuse in 94~{\AA}. As the eruption progresses and the twisted filament structure rises higher, the overlying arcade, visible as a diffuse structure in 171 and 193~{\AA} rises higher owing to the pressure created by the flux rope and the filament. AIA 131~{\AA} records emission due to \ion{Fe}{8} and \ion{Fe}{23} transitions corresponding to $\sim$0.4~MK and 14~MK, respectively \citep{2010A&A...521A..21O}. Because the overlying arcade structure is seen in 94, 171, 193 and 335 and not in 131~{\AA}, the ions in the arcade are expected to measure between 0.6~MK and 2.5~MK. 

\section{Temperature Structure of the Flux Rope} \label{temp}

Precise measurement of temperature of the observed flux rope is difficult due to the contamination of the images recorded by AIA as pointed out in Section~\ref{obs}. Therefore, it is important to estimate, if possible, the amount of the cooler emission in the hot AIA channels before anything is said about the presence of the hotter component. In order to remove the contribution due to cooler plasma from the 94~{\AA} and obtain the contribution due to \ion{Fe}{18} line, we can use the algorithms provided by \citet{2012ApJ...759..141W} or \citet{2013A&A...558A..73D}. However, this will not be sufficient as the \ion{Fe}{18} line has a rather broad contribution function and it can form anywhere between 3 to 12 MK with peak formation temperature at 7~MK. Therefore, a DEM inversion shall also be necessary to pin-point the temperature of the flux rope.

We apply the algorithm provided by \citet{2012ApJ...759..141W} for our observations of the erupting filament in the 94~{\AA} channel in order to detect the high temperature structure, if any, of the erupting filament. The eruption begins at 07:11 UT and ends around 07:35 UT. AIA 171 and 193~{\AA} channels provide an estimate of the fluxes whose ratio is used to eliminate the emission due to lower temperatures from the 94~{\AA} channel. This is done on a pixel by pixel basis. All images of the AIA used for this study were processed using standard SolarSoftware (ssw) procedures for AIA and de-rotated to the same time so as to align them perfectly. Several images were blinked to observe the accuracy of alignment of the features. These steps are essential in order to reduce the errors during the pixel to pixel subtraction process of the algorithm. The images are 201 x 220 pixels square, which corresponds to the region of interest shown by the white-box in the left panel in Figure~\ref{context}.

\begin{figure}
\begin{center}
\includegraphics[scale=0.8,trim={0.8cm 9.9cm 1cm 0.2cm},clip=true]{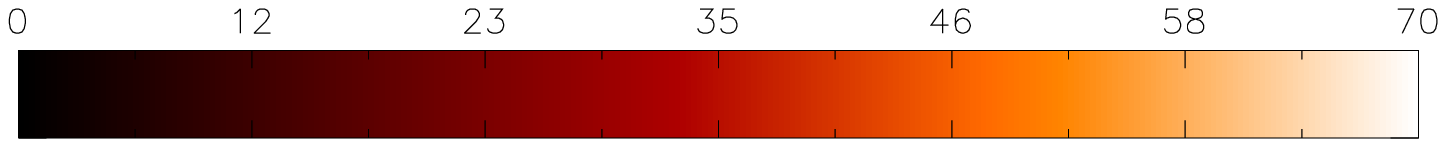}
\end{center}

\includegraphics[scale=0.6,trim={4.7cm 1.4cm 4.7cm 1cm},clip=true]{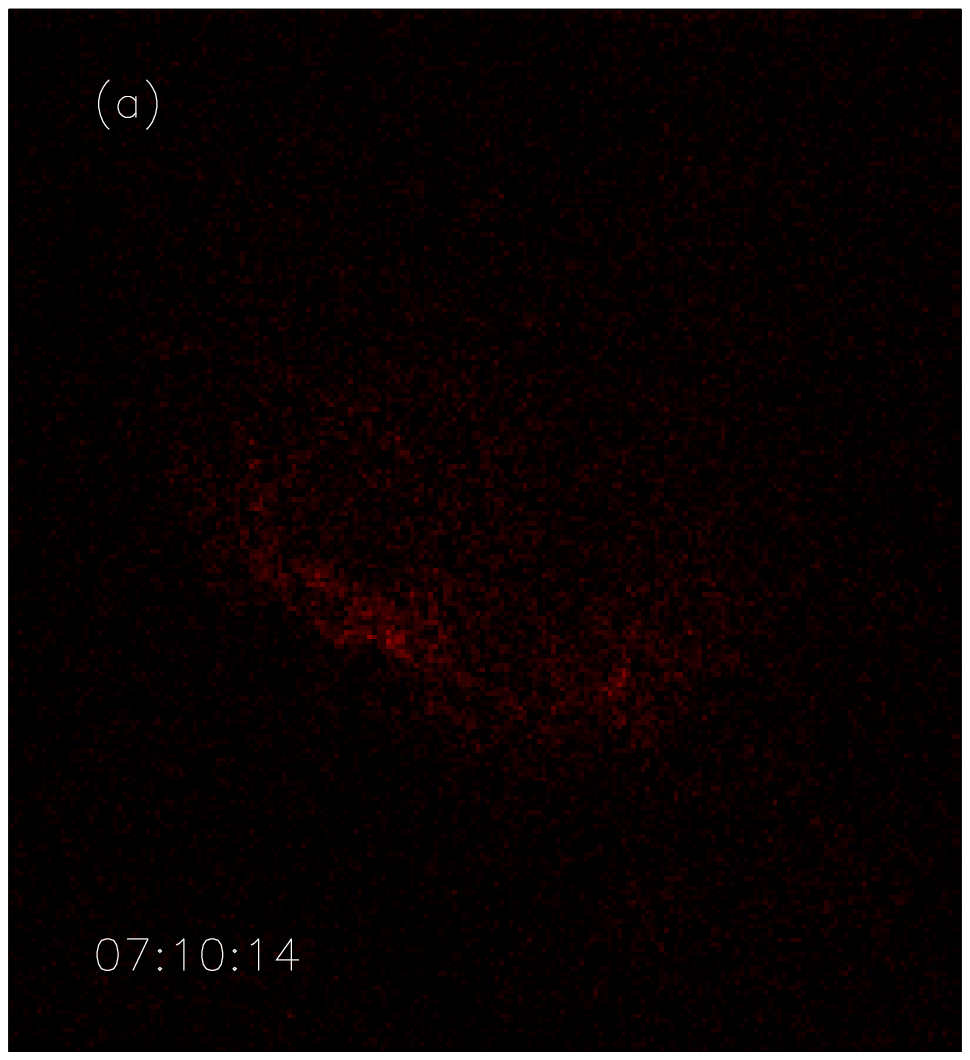}\includegraphics[scale=0.6,trim={4.7cm 1.4cm 4.7cm 1cm},clip=true]{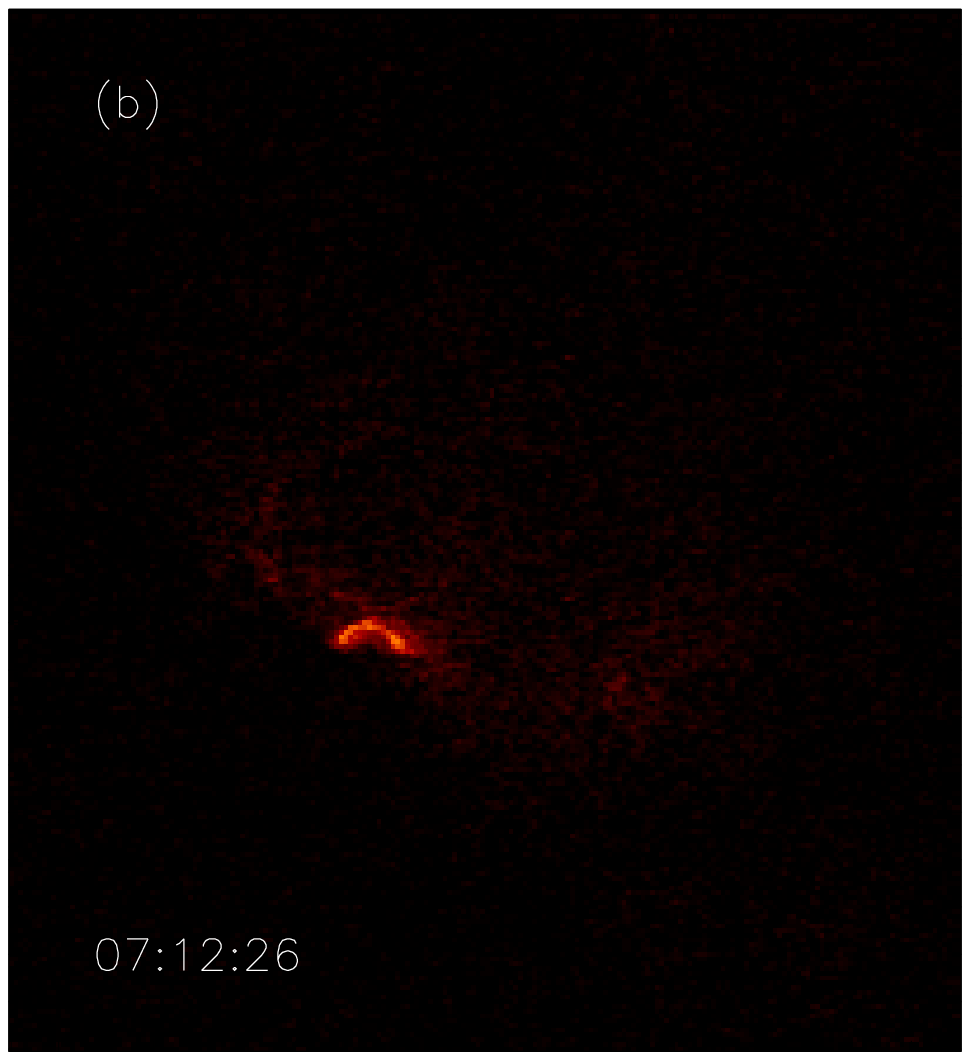}\includegraphics[scale=0.6,trim={4.7cm 1.4cm 4.7cm 1cm},clip=true]{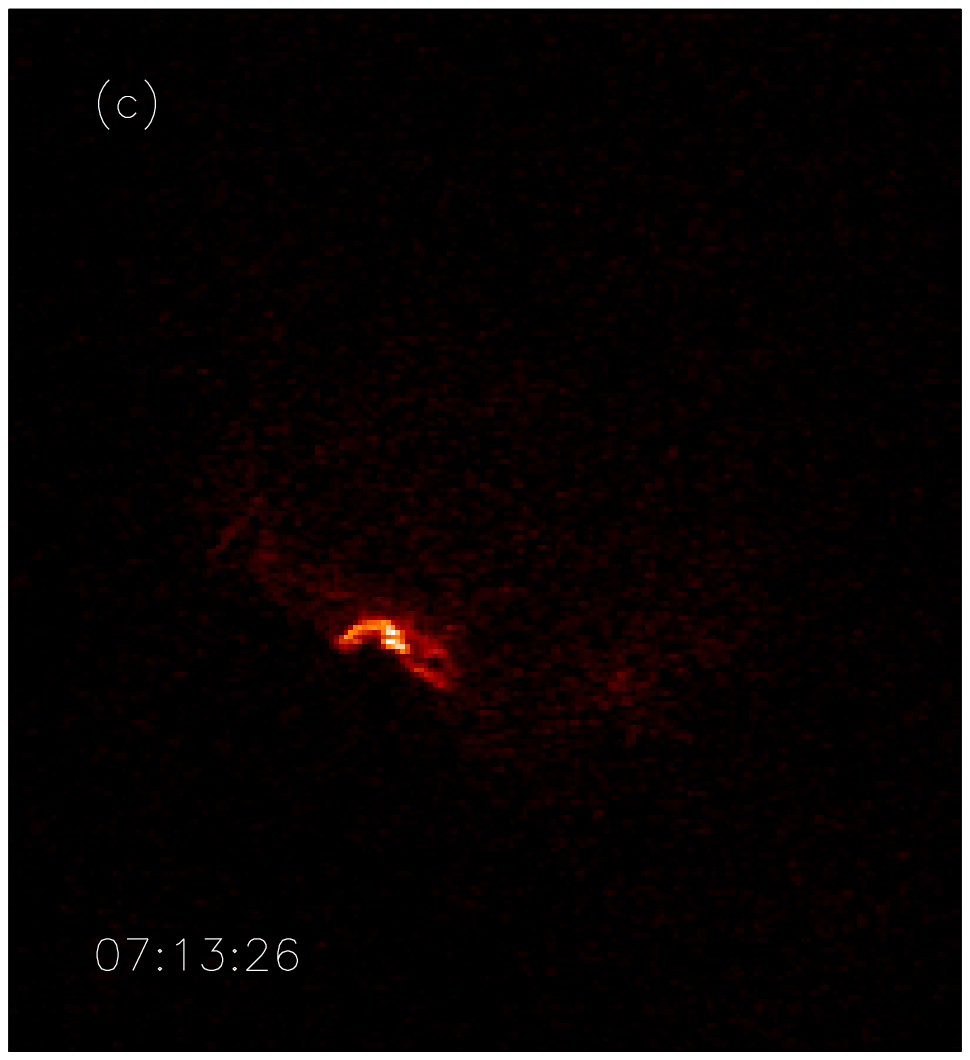}

\includegraphics[scale=0.6,trim={4.7cm 1.4cm 4.7cm 1cm},clip=true]{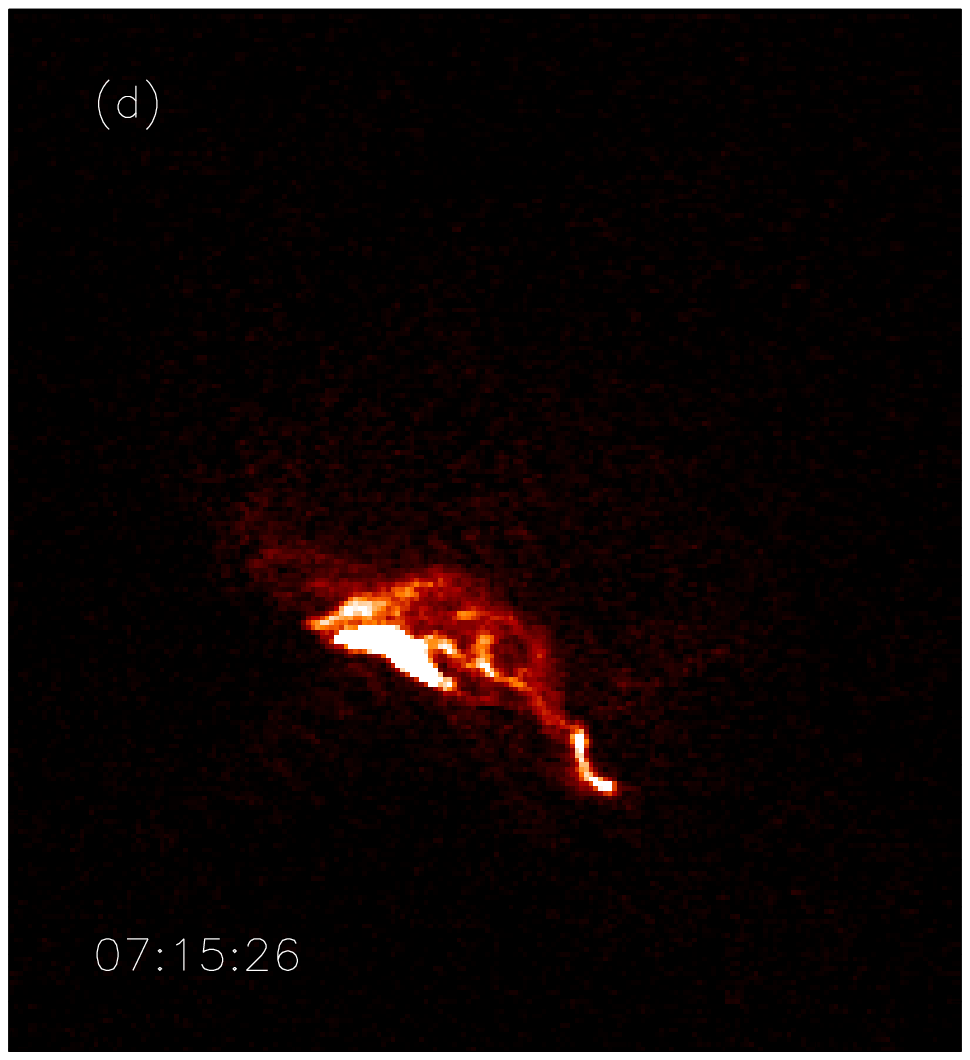}\includegraphics[scale=0.6,trim={4.7cm 1.4cm 4.7cm 1cm},clip=true]{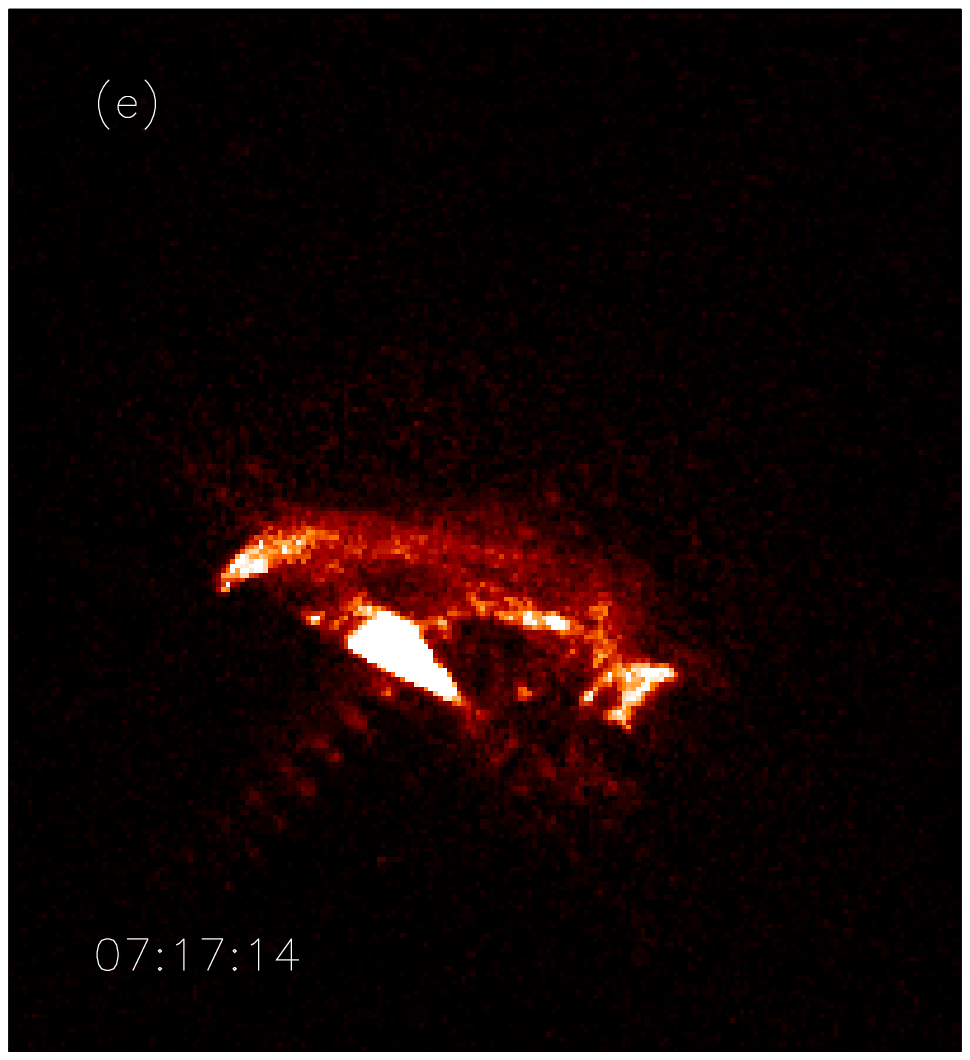}\includegraphics[scale=0.6,trim={4.7cm 1.4cm 4.7cm 1cm},clip=true]{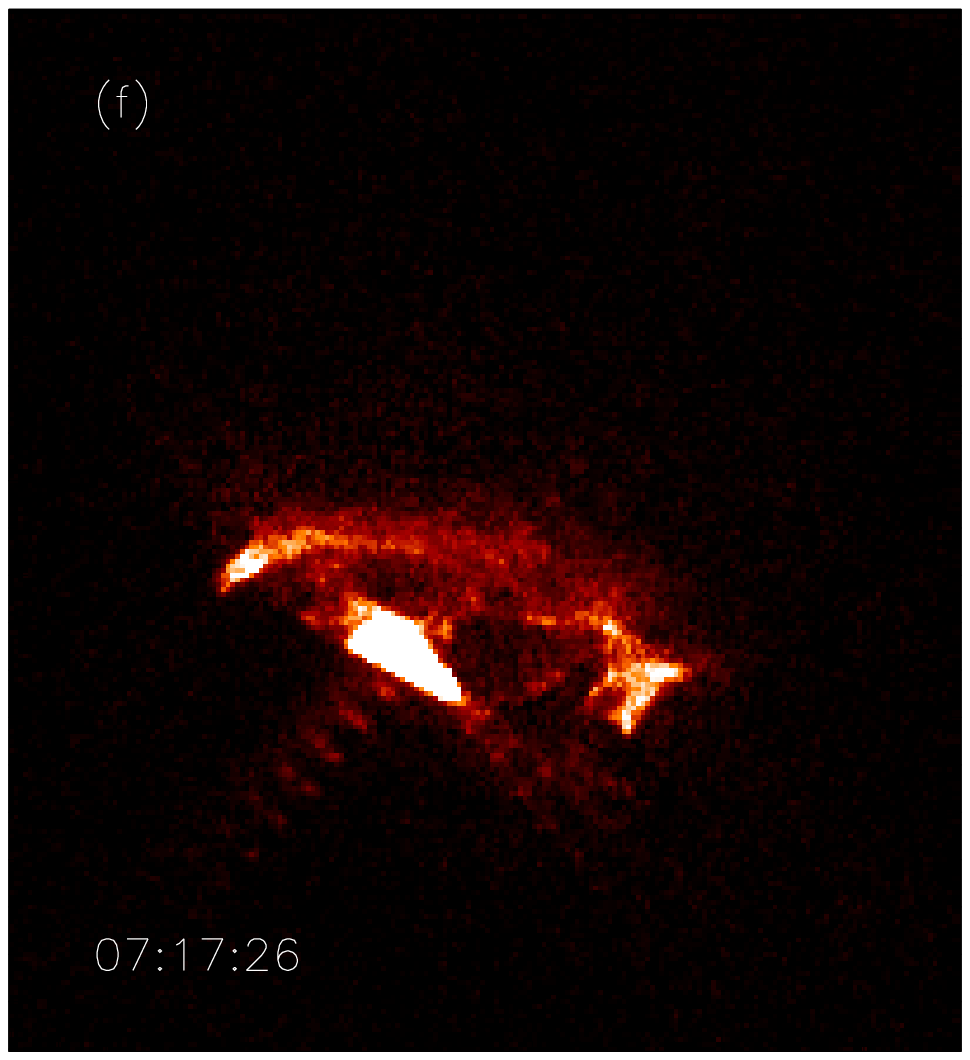}

\includegraphics[scale=0.6,trim={4.7cm 1.4cm 4.7cm 1cm},clip=true]{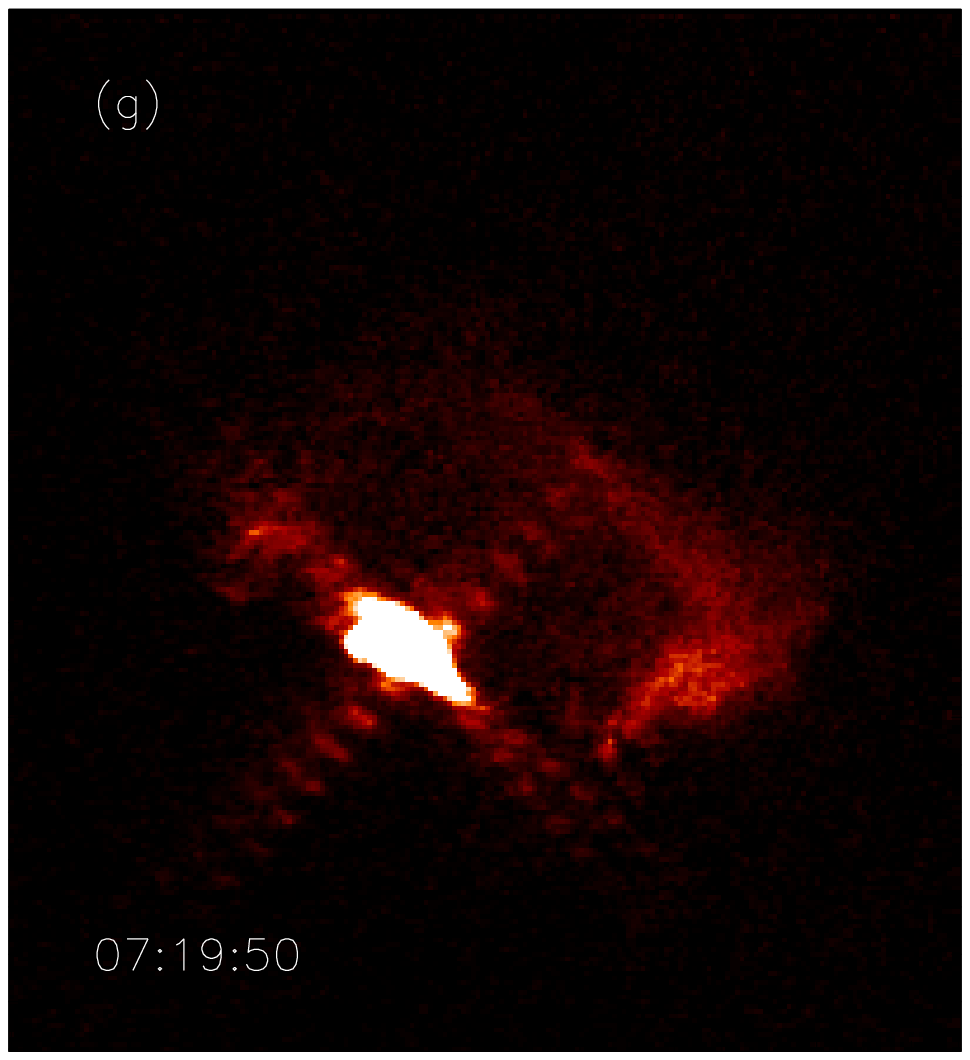}\includegraphics[scale=0.6,trim={4.7cm 1.4cm 4.7cm 1cm},clip=true]{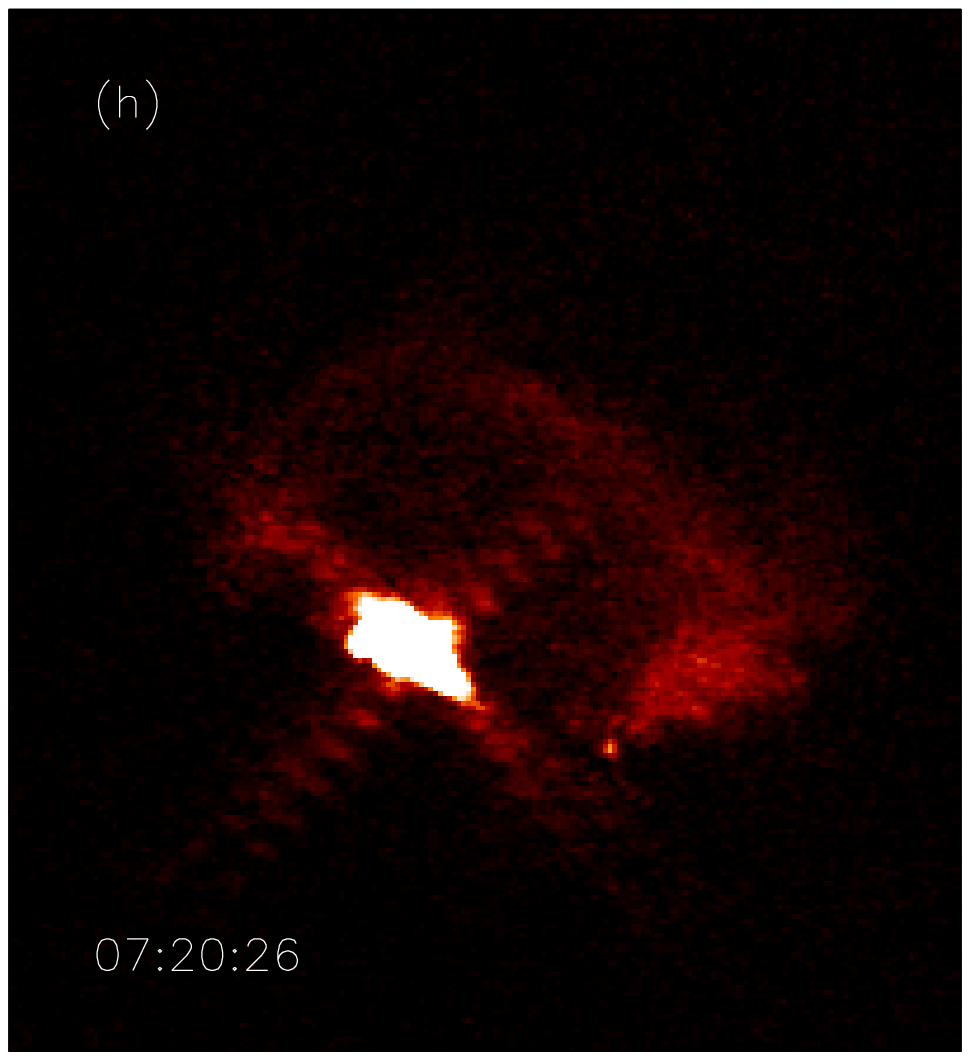}\includegraphics[scale=0.6,trim={4.7cm 1.4cm 4.7cm 1cm},clip=true]{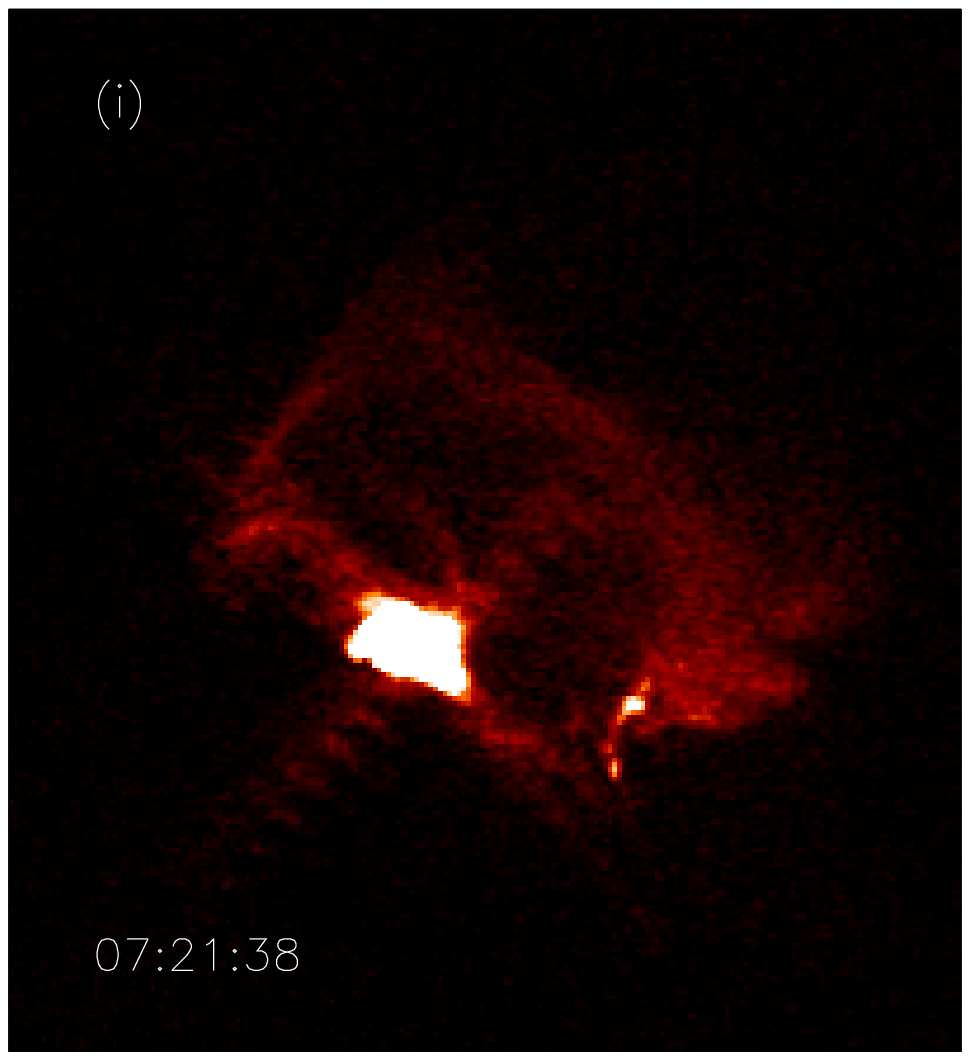}

\caption{Sequence of images taken using 94~{\AA} channel after the subtraction of cooler emission, showing the evolution of the flux rope structure at 7~MK (\ion{Fe}{18}). The intensities are in DN/s. An animation is available in the online version of this paper.\label{Fe18}}
\end{figure}

Figure~\ref{Fe18} displays a sequence of images taken using 94~{\AA} channel after the subtraction of the emission from cooler plasma. As can be seen there are no pure \ion{Fe}{18} plasma in panel (a) prior to the beginning of the eruption. Figures~\ref{Fe18}b~\&~c show small bright loops which form before the eruption begins. When observed in the \ion{Fe}{18} animation, the rise phase of the filament occurs at 07:14:14 in \ion{Fe}{18} images whereas without any subtraction, it appears to begin at 07:11:14 UT in the AIA images. This is speculated to be due to the bright emission in the reconnection region leading us to believe that the filament rise starts at an earlier time. At 07:15:26, the strands adjacent to the eastern foot of the rising filament brightens, followed by the emergence of another thread at 07:15:38 UT from the far east end (Figure\ref{Fe18}~d). The two threads continue to rise until 07:17:14~UT after which they expand as a single structure seen as the large loop in panel~(e). However, they are only partly connected to each other, some threads have ends behind the '171 loops'. That they are separate loops is evident as there is minimal emission in the region between the two from 07:18:50 and 07:19:50 (see panel g) after which some of the threads from the two loops reconnect causing an increase in the emission and a continuous structure (panel i). As the filament rises, the threads get tether-cut (panel f) after which the filament structures seen in cooler AIA channels are not seen in \ion{Fe}{18} images. This implies that the filament threads have cooled to a lower temperature and thus not visible in the \ion{Fe}{18}.

We believe that the expanding loop structure in the \ion{Fe}{18} images is the associated flux rope that is initially twisted at the location where the small loops are seen in the \ion{Fe}{18} images. The small loops form due to tether-cutting reconnection and the filament threads are cut during the reconnection due to which it erupts followed by the expansion of the flux rope. The image sequence in Figure~\ref{Fe18} reveals the hot nature of the flux rope as the cooler component from the 94~{\AA} channel has been removed. For a few 94 ~{\AA} images during the course of the eruption, we compared the results obtained using \citeauthor{2012ApJ...759..141W} algorithm with that obtained using \citet{2013A&A...558A..73D}'s algorithm  and found that the results were generally consistent with each other. 

\begin{figure}\label{emreg}
\includegraphics[scale=0.6,trim={1cm 3cm 1cm 3cm},clip=true]{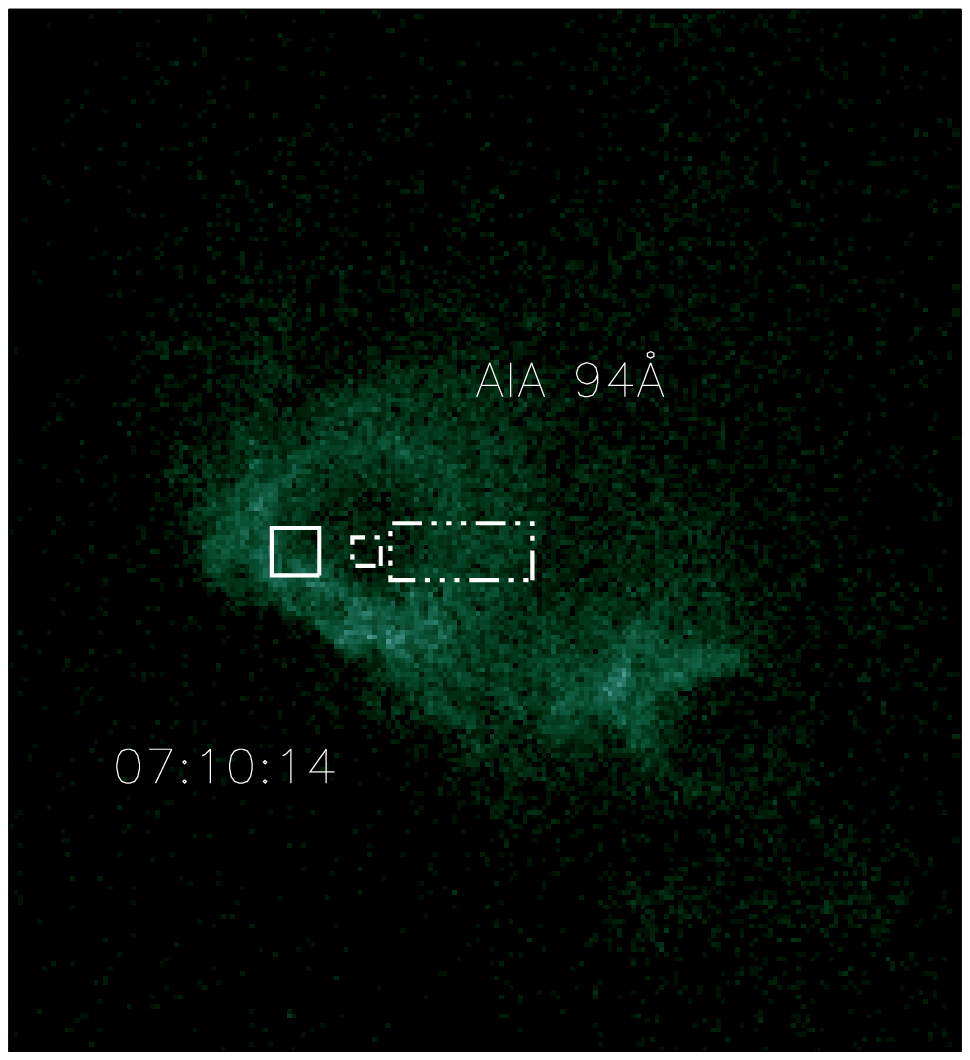}\includegraphics[scale=0.6,trim={1cm 3cm 1cm 3cm},clip=true]{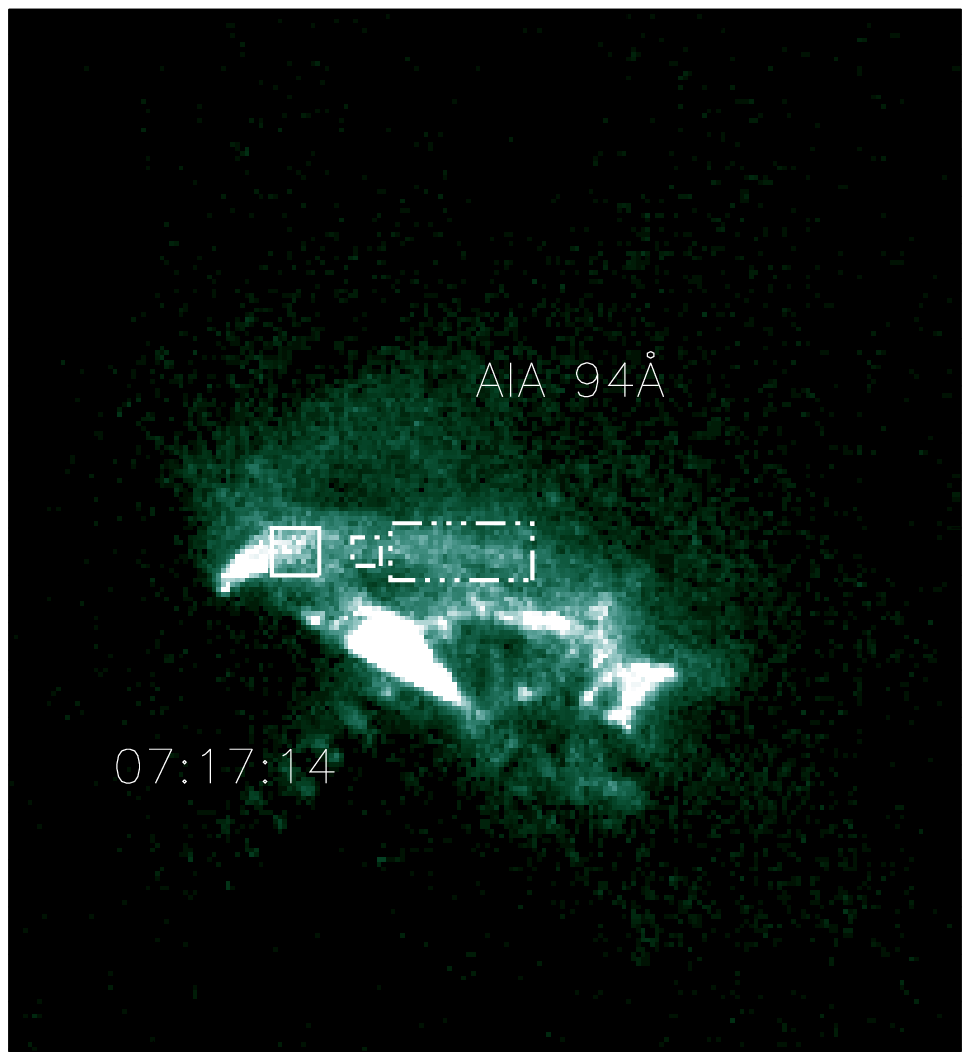}

\includegraphics[scale=0.42,trim={1.2cm 0.4cm 0.8cm 0.2cm},clip=true]{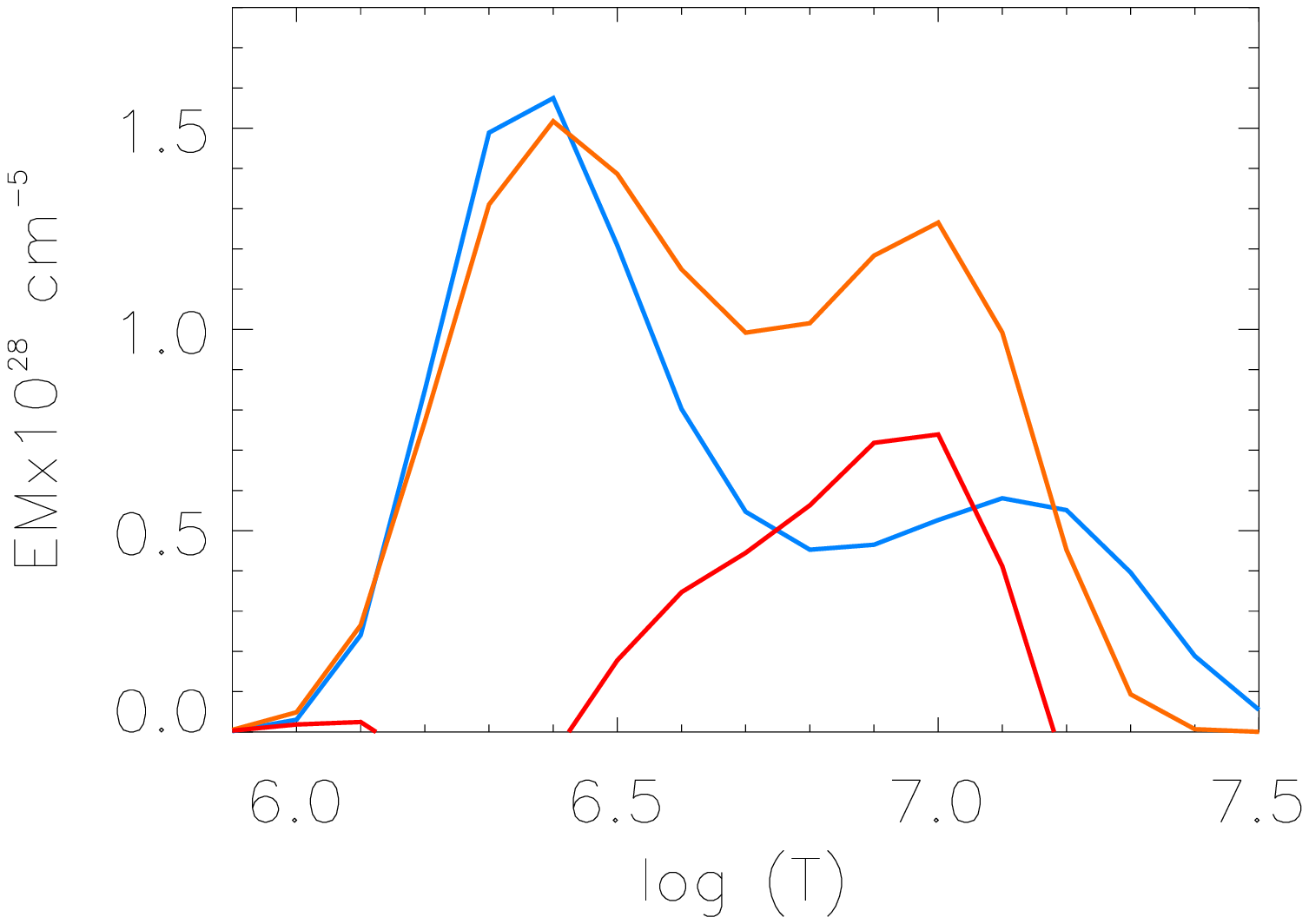}\includegraphics[scale=0.42,trim={3.3cm 0.4cm 0.8cm 0.2cm}, clip=true]{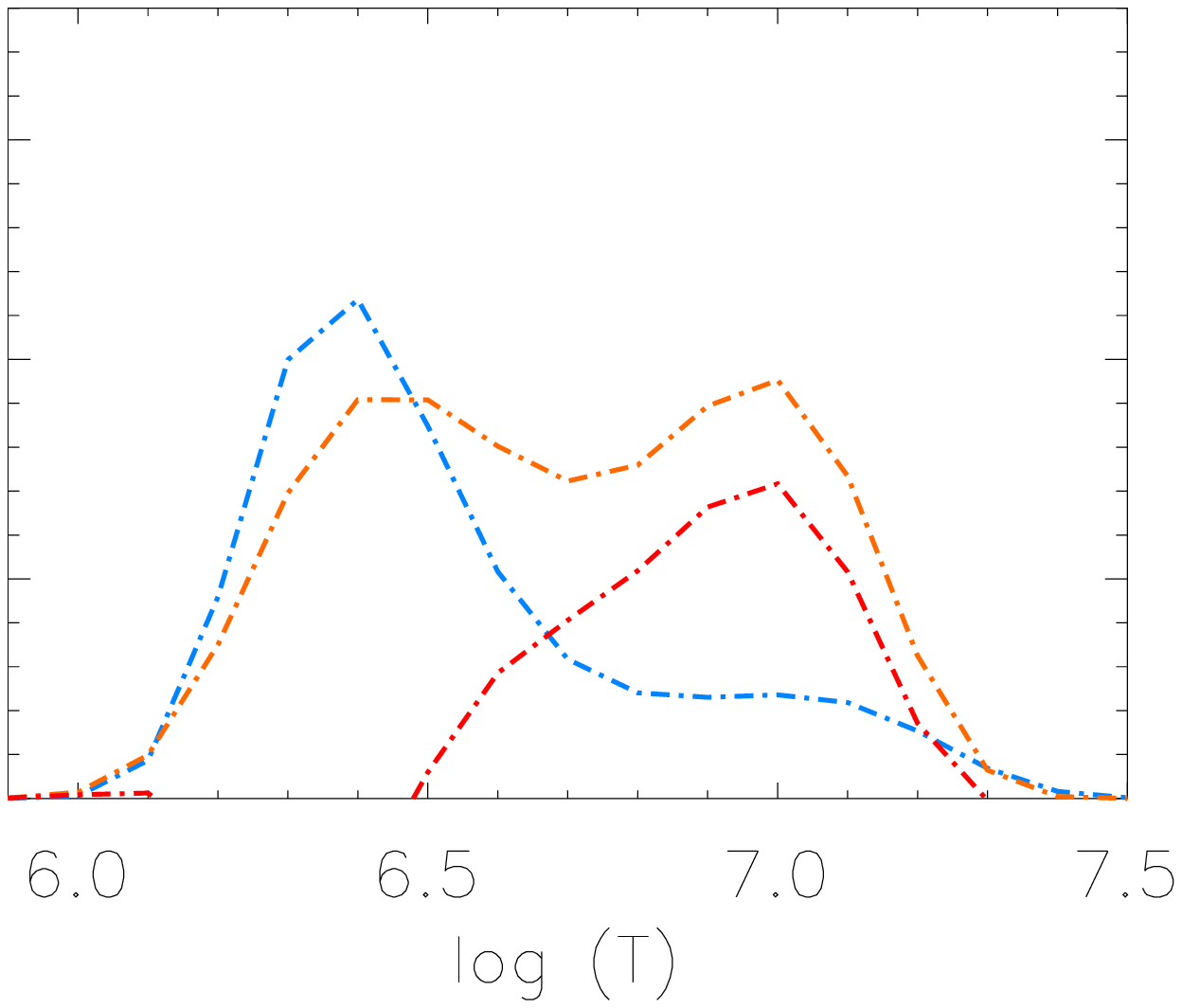}\includegraphics[scale=0.42,trim={3.3cm 0.4cm 0.8cm 0.2cm},clip=true]{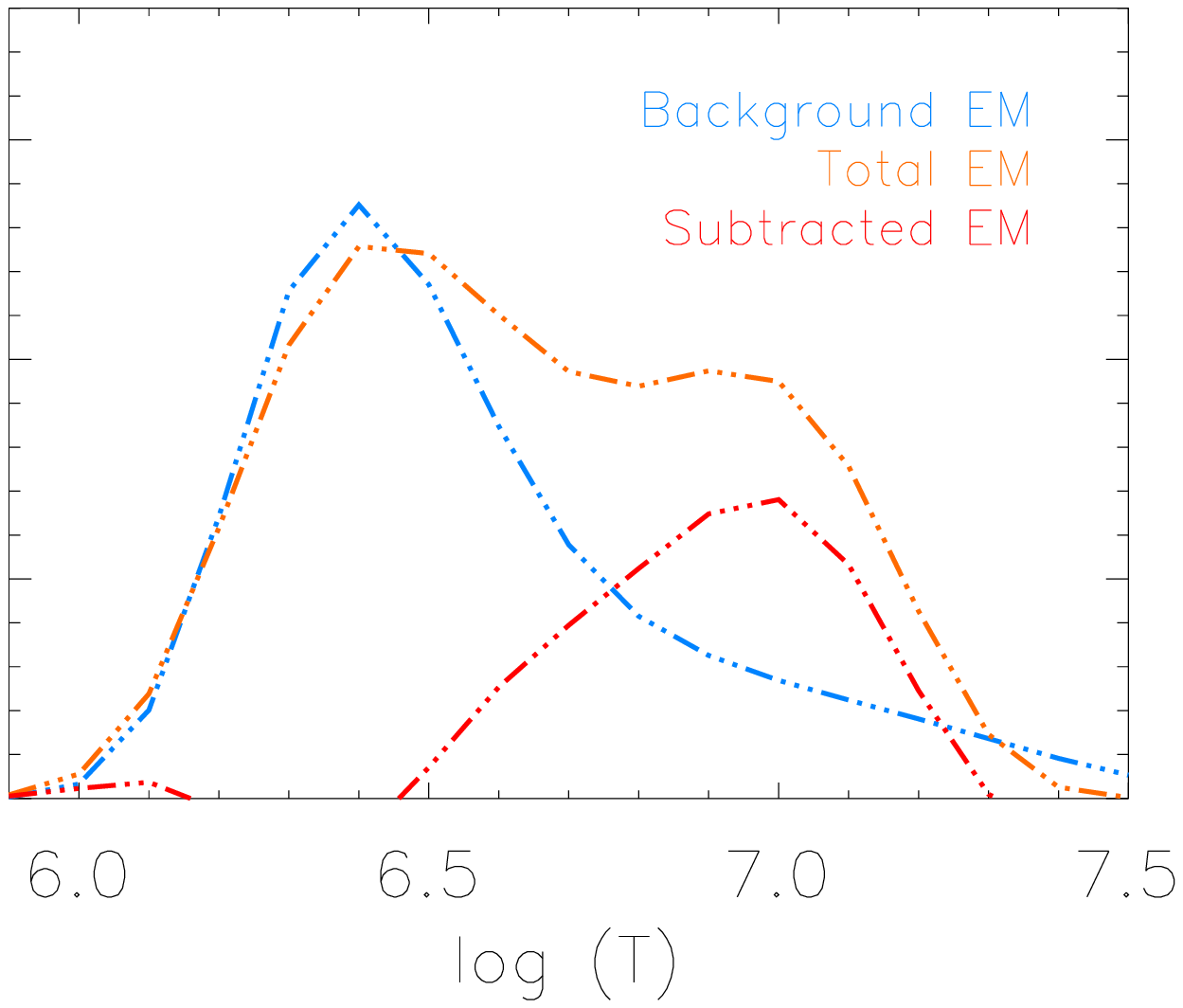}
\caption{The top panel shows the regions selected to calculate the emission measure (EM) of the flux rope (right) and the background (left). The bottom panel shows the background, total and the subtracted emission measures of the three regions. The EM is calulated using 94, 131, 171, 193, 211 \& 335~{\AA} channels. The different linestyles in the plots corresponds to the regions marked with similar linestyles in the top panel. \label{emplot}}
\end{figure}

In order to compare the temperature range of the flux rope using the DEM technique, we compute DEM of various regions of the flux rope (shown on the right in the top panel in Figure~\ref{emplot}). For this purpose we have adopted the IDL routine provided in SolarSoft \textit{xrt\_dem\_iterative2.pro} \citep{2004IAUS..223..321W} that computes column differential emission measure. Pre-eruption DEM of the same region (shown on the left in the top panel of Figure~\ref{emplot}) is also computed in order to remove the emissions due to the background and foreground. For this computation, we have used Chianti v8 \citep{1997A&AS..125..149D,2015A&A...582A..56D} with abundances from \citet{1998SSRv...85..161G}. The EM obtained from the calculated DEM for the three regions, the background and the subtracted EM are shown in the bottom panel in Figure~\ref{emplot}. As can be seen, the background EM (i.e. EM computed before the start of the eruption) peaks at $\log\ T = 6.5$, the total EM (i.e. at the time when the flux rope was clearly visible in \ion{Fe}{18} images) peaks at around $\log\ T = 6.8$. We note that the low temperature peak is very closely matched with the peak for the background region, suggesting that the cooler temperature peak in the EM curve of the flux rope is most likely due to the omnipresent background emission \citep{2014ApJ...795...76S} and may not have anything to do with the erupting flux rope structure. The second peak of the Background EM shown using the smooth line-style is attributed to the heated background as seen in the 94A image on the left in Figure~\ref{emplot}. The red curve is the difference of the two EM curves, peaking at $\log\ T = 7.0$, providing a range of temperature between $\log\ T = 6.7$ and $\log\ T = 7.2$, which roughly coincides with the contribution function of \ion{Fe}{18} line formed at 93.9~{\AA}.

\section{Summary} \label{discussion}

An erupting filament observed by SDO/AIA on Oct 31, 2010, between 07:11 and 07:35 UT in all the EUV channels is studied. We have focussed on the kinematics of the eruption using six EUV channels, identified a flux rope associated with the filament and obtained its temperature. The main findings of the study are as described below: 

\begin{enumerate}
	
\item The filament shows slow rise followed by a fast rise phase. Local brightening at the location from where the eruption initially begins is seen following which the slow-rise phase begins. The eruption is asymmetric and failed. Towards the end of the eruption, the filament material drains towards the solar surface along twisted field lines of one leg of the filament.

\item At the location of the brightening from where the filament eruption occurs, loops are seen when the pixel saturation levels drop towards the end of the eruption. We believe that tether-cutting type reconnection mechanism is causing the filament eruption and the formation of the loops.

\item The flux rope associated with the filament starts to rise about two minutes after the filament eruptions begins. It is seen only in the hot channels - 94 and 131~{\AA} ($\sim 11MK$) images of AIA. The twisted structure of the flux rope is also identified. 

\item Since the AIA channels have contributions from other cooler lines and a definitive temperature measure have not been established on flux ropes, we use an algorithm devised by \citet{2012ApJ...759..141W} to remove the cooler emission component. This is used to study the thermal structure and evolution  of the flux rope with time. During the initial slow rise phase, the flux rope is visible both in the \ion{Fe}{18} and the 131~{\AA} images, but in the later phase, the structure is visible only in \ion{Fe}{18}.

\item DEM analysis at one instant of time is performed to compare the temperature structure obtained using the algorithm and that using the DEM. The DEM analysis gives a range of temperature between $\log\ T = 6.7$ and $\log\ T = 7.2$ and peaks at $\log\ T = 7.2$, corresponding to the formation temperature of \ion{Fe}{18} line at 93.9~{\AA}.

\end{enumerate}

Various previous studies have identified and analysed flux ropes in hot as well as cool channels of AIA and provide a broad range of temperature for the flux ropes \citep{2011ApJ...727L..52R, 2015A&A...580A...2Z, 2015ApJ...808..117N, 2012ApJ...761...62C, 2013ApJ...778..142T}. These studies approximately provide the temperature ranges of the features during eruptions based on observations in high temperature channels of AIA and DEM analysis. The last two studies attribute the cause of emission in higher temperature to reconnection. Here, we provide with certainty the reconnection mechanism at the site of eruption causing the eruption and the formation of the  loops. 

The results obtained in this study lead us to conclude that the flux ropes are hot structures with temperatures of about 11~MK in the early phase of the eruption, which cools down to 7~MK during the evolution, very likely due to expansion. Further studies are required to make this constraint more front footing. The spontaneous formation of the flux rope is not observed, also the presence of the flux rope is not detected prior to the eruption begins. However, the tether-cutting reconnection at the location from where the eruption begins suggests the formation of a flux rope or create additional twists in an already existing flux rope due to reconnection. The isolation of the cool components of emission provides a measure to detect the hot flux ropes. Constraints on the temperature of the flux ropes will further be useful for thermodynamic modelling of flux ropes in CMEs. 

\acknowledgments{We thank the referee for constructive comments that has improved the manuscript. The authors acknowledge the support from DST under the Fast Track Scheme (SERB/F/3369/2012-2013). The AIA data are a courtesy of SDO (NASA) and the AIA consortium. CHIANTI is a collaborative project involving George Mason University, the University of Michigan (USA) and the University of Cambridge (UK). The authors thank Helen Mason for useful discussions.}


\begin{thebibliography}{35}
\expandafter\ifx\csname natexlab\endcsname\relax\def\natexlab#1{#1}\fi

\bibitem[{{Boerner} {et~al.}(2012){Boerner}, {Edwards}, {Lemen}, {Rausch},
  {Schrijver}, {Shine}, {Shing}, {Stern}, {Tarbell}, {Title}, {Wolfson},
  {Soufli}, {Spiller}, {Gullikson}, {McKenzie}, {Windt}, {Golub}, {Podgorski},
  {Testa}, \& {Weber}}]{2012SoPh..275...41B}
{Boerner}, P., {et~al.} 2012, \solphys, 275, 41

\bibitem[{{Brueckner} {et~al.}(1995){Brueckner}, {Howard}, {Koomen},
  {Korendyke}, {Michels}, {Moses}, {Socker}, {Dere}, {Lamy}, {Llebaria},
  {Bout}, {Schwenn}, {Simnett}, {Bedford}, \& {Eyles}}]{1995SoPh..162..357B}
{Brueckner}, G.~E., {et~al.} 1995, \solphys, 162, 357

\bibitem[{{Cecconi} {et~al.}(2008){Cecconi}, {Bonnin}, {Hoang}, {Maksimovic},
  {Bale}, {Bougeret}, {Goetz}, {Lecacheux}, {Reiner}, {Rucker}, \&
  {Zarka}}]{2008SSRv..136..549C}
{Cecconi}, B., {et~al.} 2008, \ssr, 136, 549

\bibitem[{{Cheng} {et~al.}(2012){Cheng}, {Zhang}, {Saar}, \&
  {Ding}}]{2012ApJ...761...62C}
{Cheng}, X., {Zhang}, J., {Saar}, S.~H., \& {Ding}, M.~D. 2012, \apj, 761, 62

\bibitem[{{Chifor} {et~al.}(2006){Chifor}, {Mason}, {Tripathi}, {Isobe}, \&
  {Asai}}]{2006A&A...458..965C}
{Chifor}, C., {Mason}, H.~E., {Tripathi}, D., {Isobe}, H., \& {Asai}, A. 2006,
  \aap, 458, 965

\bibitem[{{Chifor} {et~al.}(2007){Chifor}, {Tripathi}, {Mason}, \&
  {Dennis}}]{2007A&A...472..967C}
{Chifor}, C., {Tripathi}, D., {Mason}, H.~E., \& {Dennis}, B.~R. 2007, \aap,
  472, 967

\bibitem[{{Cremades} \& {Bothmer}(2004)}]{2004A&A...422..307C}
{Cremades}, H., \& {Bothmer}, V. 2004, \aap, 422, 307

\bibitem[{{Del Zanna}(2013)}]{2013A&A...558A..73D}
{Del Zanna}, G. 2013, \aap, 558, A73

\bibitem[{{Del Zanna} {et~al.}(2015){Del Zanna}, {Dere}, {Young}, {Landi}, \&
  {Mason}}]{2015A&A...582A..56D}
{Del Zanna}, G., {Dere}, K.~P., {Young}, P.~R., {Landi}, E., \& {Mason}, H.~E.
  2015, \aap, 582, A56

\bibitem[{{Del Zanna} {et~al.}(2011){Del Zanna}, {O'Dwyer}, \&
  {Mason}}]{2011A&A...535A..46D}
{Del Zanna}, G., {O'Dwyer}, B., \& {Mason}, H.~E. 2011, \aap, 535, A46

\bibitem[{{Dere} {et~al.}(1997){Dere}, {Landi}, {Mason}, {Monsignori Fossi}, \&
  {Young}}]{1997A&AS..125..149D}
{Dere}, K.~P., {Landi}, E., {Mason}, H.~E., {Monsignori Fossi}, B.~C., \&
  {Young}, P.~R. 1997, \aaps, 125, 149

\bibitem[{{Gibson} \& {Fan}(2006)}]{2006JGRA..11112103G}
{Gibson}, S.~E., \& {Fan}, Y. 2006, Journal of Geophysical Research (Space
  Physics), 111, 12103

\bibitem[{{Gibson} {et~al.}(2006){Gibson}, {Fan}, {T{\"o}r{\"o}k}, \&
  {Kliem}}]{2006SSRv..124..131G}
{Gibson}, S.~E., {Fan}, Y., {T{\"o}r{\"o}k}, T., \& {Kliem}, B. 2006, \ssr,
  124, 131

\bibitem[{{Gibson} {et~al.}(2010){Gibson}, {Kucera}, {Rastawicki}, {Dove}, {de
  Toma}, {Hao}, {Hill}, {Hudson}, {Marqu{\'e}}, {McIntosh}, {Rachmeler},
  {Reeves}, {Schmieder}, {Schmit}, {Seaton}, {Sterling}, {Tripathi},
  {Williams}, \& {Zhang}}]{2010ApJ...724.1133G}
{Gibson}, S.~E., {et~al.} 2010, \apj, 724, 1133

\bibitem[{{Grevesse} \& {Sauval}(1998)}]{1998SSRv...85..161G}
{Grevesse}, N., \& {Sauval}, A.~J. 1998, \ssr, 85, 161

\bibitem[{{Howard} {et~al.}(2008){Howard}, {Moses}, {Vourlidas}, {Newmark},
  {Socker}, {Plunkett}, {Korendyke}, {Cook}, {Hurley}, {Davila}, {Thompson},
  {St Cyr}, {Mentzell}, {Mehalick}, {Lemen}, {Wuelser}, {Duncan}, {Tarbell},
  {Wolfson}, {Moore}, {Harrison}, {Waltham}, {Lang}, {Davis}, {Eyles},
  {Mapson-Menard}, {Simnett}, {Halain}, {Defise}, {Mazy}, {Rochus}, {Mercier},
  {Ravet}, {Delmotte}, {Auchere}, {Delaboudiniere}, {Bothmer}, {Deutsch},
  {Wang}, {Rich}, {Cooper}, {Stephens}, {Maahs}, {Baugh}, {McMullin}, \&
  {Carter}}]{2008SSRv..136...67H}
{Howard}, R.~A., {et~al.} 2008, \ssr, 136, 67

\bibitem[{{Kucera} {et~al.}(2012){Kucera}, {Gibson}, {Schmit}, {Landi}, \&
  {Tripathi}}]{2012ApJ...757...73K}
{Kucera}, T.~A., {Gibson}, S.~E., {Schmit}, D.~J., {Landi}, E., \& {Tripathi},
  D. 2012, \apj, 757, 73

\bibitem[{{Lemen} {et~al.}(2012){Lemen}, {Title}, {Akin}, {Boerner}, {Chou},
  {Drake}, {Duncan}, {Edwards}, {Friedlaender}, {Heyman}, {Hurlburt}, {Katz},
  {Kushner}, {Levay}, {Lindgren}, {Mathur}, {McFeaters}, {Mitchell}, {Rehse},
  {Schrijver}, {Springer}, {Stern}, {Tarbell}, {Wuelser}, {Wolfson}, {Yanari},
  {Bookbinder}, {Cheimets}, {Caldwell}, {Deluca}, {Gates}, {Golub}, {Park},
  {Podgorski}, {Bush}, {Scherrer}, {Gummin}, {Smith}, {Auker}, {Jerram},
  {Pool}, {Soufli}, {Windt}, {Beardsley}, {Clapp}, {Lang}, \&
  {Waltham}}]{2012SoPh..275...17L}
{Lemen}, J.~R., {et~al.} 2012, \solphys, 275, 17

\bibitem[{{Liu} {et~al.}(2013){Liu}, {Deng}, {Lee}, {Wiegelmann}, {Moore}, \&
  {Wang}}]{2013ApJ...778L..36L}
{Liu}, C., {Deng}, N., {Lee}, J., {Wiegelmann}, T., {Moore}, R.~L., \& {Wang},
  H. 2013, \apjl, 778, L36

\bibitem[{{Mackay} {et~al.}(2010){Mackay}, {Karpen}, {Ballester}, {Schmieder},
  \& {Aulanier}}]{2010SSRv..151..333M}
{Mackay}, D.~H., {Karpen}, J.~T., {Ballester}, J.~L., {Schmieder}, B., \&
  {Aulanier}, G. 2010, \ssr, 151, 333

\bibitem[{{Moore} {et~al.}(2001){Moore}, {Sterling}, {Hudson}, \&
  {Lemen}}]{2001ApJ...552..833M}
{Moore}, R.~L., {Sterling}, A.~C., {Hudson}, H.~S., \& {Lemen}, J.~R. 2001,
  \apj, 552, 833

\bibitem[{{Nindos} {et~al.}(2015){Nindos}, {Patsourakos}, {Vourlidas}, \&
  {Tagikas}}]{2015ApJ...808..117N}
{Nindos}, A., {Patsourakos}, S., {Vourlidas}, A., \& {Tagikas}, C. 2015, \apj,
  808, 117

\bibitem[{{O'Dwyer} {et~al.}(2010){O'Dwyer}, {Del Zanna}, {Mason}, {Weber}, \&
  {Tripathi}}]{2010A&A...521A..21O}
{O'Dwyer}, B., {Del Zanna}, G., {Mason}, H.~E., {Weber}, M.~A., \& {Tripathi},
  D. 2010, \aap, 521, A21

\bibitem[{{Patsourakos} {et~al.}(2013){Patsourakos}, {Vourlidas}, \&
  {Stenborg}}]{2013ApJ...764..125P}
{Patsourakos}, S., {Vourlidas}, A., \& {Stenborg}, G. 2013, \apj, 764, 125

\bibitem[{{Reeves} \& {Golub}(2011)}]{2011ApJ...727L..52R}
{Reeves}, K.~K., \& {Golub}, L. 2011, \apjl, 727, L52

\bibitem[{{Schmit} {et~al.}(2009){Schmit}, {Gibson}, {Tomczyk}, {Reeves},
  {Sterling}, {Brooks}, {Williams}, \& {Tripathi}}]{2009ApJ...700L..96S}
{Schmit}, D.~J., {Gibson}, S.~E., {Tomczyk}, S., {Reeves}, K.~K., {Sterling},
  A.~C., {Brooks}, D.~H., {Williams}, D.~R., \& {Tripathi}, D. 2009, \apjl,
  700, L96

\bibitem[{{Sheeley} {et~al.}(2014){Sheeley}, {Warren}, {Lee}, {Chung}, {Katz},
  \& {Namkung}}]{2014ApJ...797..131S}
{Sheeley}, Jr., N.~R., {Warren}, H.~P., {Lee}, J., {Chung}, S., {Katz}, J., \&
  {Namkung}, M. 2014, \apj, 797, 131

\bibitem[{{Subramanian} {et~al.}(2014){Subramanian}, {Tripathi}, {Klimchuk}, \&
  {Mason}}]{2014ApJ...795...76S}
{Subramanian}, S., {Tripathi}, D., {Klimchuk}, J.~A., \& {Mason}, H.~E. 2014,
  \apj, 795, 76

\bibitem[{{Tripathi} {et~al.}(2009{\natexlab{a}}){Tripathi}, {Gibson}, {Qiu},
  {Fletcher}, {Liu}, {Gilbert}, \& {Mason}}]{2009A&A...498..295T}
{Tripathi}, D., {Gibson}, S.~E., {Qiu}, J., {Fletcher}, L., {Liu}, R.,
  {Gilbert}, H., \& {Mason}, H.~E. 2009{\natexlab{a}}, \aap, 498, 295

\bibitem[{{Tripathi} {et~al.}(2006){Tripathi}, {Isobe}, \&
  {Mason}}]{2006A&A...453.1111T}
{Tripathi}, D., {Isobe}, H., \& {Mason}, H.~E. 2006, \aap, 453, 1111

\bibitem[{{Tripathi} {et~al.}(2009{\natexlab{b}}){Tripathi}, {Kliem}, {Mason},
  {Young}, \& {Green}}]{2009ApJ...698L..27T}
{Tripathi}, D., {Kliem}, B., {Mason}, H.~E., {Young}, P.~R., \& {Green}, L.~M.
  2009{\natexlab{b}}, \apjl, 698, L27

\bibitem[{{Tripathi} {et~al.}(2013){Tripathi}, {Reeves}, {Gibson},
  {Srivastava}, \& {Joshi}}]{2013ApJ...778..142T}
{Tripathi}, D., {Reeves}, K.~K., {Gibson}, S.~E., {Srivastava}, A., \& {Joshi},
  N.~C. 2013, \apj, 778, 142

\bibitem[{{Warren} {et~al.}(2012){Warren}, {Winebarger}, \&
  {Brooks}}]{2012ApJ...759..141W}
{Warren}, H.~P., {Winebarger}, A.~R., \& {Brooks}, D.~H. 2012, \apj, 759, 141

\bibitem[{{Weber} {et~al.}(2004){Weber}, {Deluca}, {Golub}, \&
  {Sette}}]{2004IAUS..223..321W}
{Weber}, M.~A., {Deluca}, E.~E., {Golub}, L., \& {Sette}, A.~L. 2004, in IAU
  Symposium, Vol. 223, Multi-Wavelength Investigations of Solar Activity, ed.
  A.~V. {Stepanov}, E.~E. {Benevolenskaya}, \& A.~G. {Kosovichev}, 321--328

\bibitem[{{Zhang} {et~al.}(2015){Zhang}, {Yang}, \& {Li}}]{2015A&A...580A...2Z}
{Zhang}, J., {Yang}, S.~H., \& {Li}, T. 2015, \aap, 580, A2

\end{thebibliography}

\end{document}